\documentclass[sigconf,screen]{acmart}

\copyrightyear{2026}
\acmYear{2026}
\setcopyright{cc}
\setcctype{by}
\acmConference[CHI '26]{Proceedings of the 2026 CHI Conference on Human Factors in Computing Systems}{April 13--17, 2026}{Barcelona, Spain}
\acmBooktitle{Proceedings of the 2026 CHI Conference on Human Factors in Computing Systems (CHI '26), April 13--17, 2026, Barcelona, Spain}
\acmDOI{}
\acmISBN{}

\usepackage{float}
\usepackage{booktabs}
\usepackage{tabularx}
\usepackage{threeparttable} 
\usepackage{placeins}
\usepackage{enumitem}
\usepackage{natbib}
\usepackage{subcaption}
\usepackage{xcolor}
\usepackage{soul}

\newcommand \revision[1]{{\textcolor{black}{#1}}}

\AtBeginDocument{%
  }\AtBeginDocument{%
  }

\begin{document}

\title{SituFont: A Just-in-Time Adaptive Intervention Interface for Enhancing Mobile Readability in Situational Visual Impairments}


\author{Jingruo Chen}
\orcid{0009-0007-1606-5780}
\email{jc3564@cornell.edu}
\affiliation{%
  \institution{Cornell University}
  \city{Ithaca}
  \state{New York}
  \country{USA}
}
\authornote{These authors contributed equally to this research.}

\author{Kexin Nie}
\orcid{0009-0002-9190-092X}
\email{niekexinbella@gmail.com}
\affiliation{%
  \institution{The University of Sydney}
  \city{Sydney}
  \country{Australia}
}
\authornotemark[1]

\author{Mingshan Zhang}
\orcid{0009-0000-2196-304X}
\affiliation{%
  \institution{Tsinghua University}
  \city{Beijing}
  \country{China}
}
\authornotemark[1]

\author{Chun Yu}
\orcid{0000-0003-2591-7993}
\affiliation{%
  \institution{Tsinghua University}
  \city{Beijing}
  \country{China}
}

\author{Zhiqi Gao}
\orcid{0009-0008-4898-2521}
\affiliation{%
  \institution{Nankai University}
  \city{Tianjin}
  \country{China}
}

\author{Kun Yue}
\orcid{0009-0009-2965-559X}
\email{kunyue@infera.cn}
\affiliation{%
  \institution{Tsinghua University}
  \city{Beijing}
  \country{China}
}

\author{Chen Liang}
\orcid{0000-0003-0579-2716}
\affiliation{%
  \institution{The Hong Kong University of Science and Technology (Guangzhou)}
  \city{Guangzhou}
  \state{Guangdong}
  \country{China}
}
\authornote{Corresponding author.}

\author{Yuanchun Shi}
\orcid{0000-0003-2273-6927}
\affiliation{%
  \institution{Tsinghua University}
  \city{Beijing}
  \country{China}
}

\renewcommand{\shortauthors }{Chen et al.}

\begin{abstract}

Situational visual impairments (SVIs) hinder mobile readability, causing discomfort and limiting information access. Building on prior work in adaptive typography and accessibility, this paper presents SituFont, a \revision{context-aware and human-in-the-loop adaptive typography adjustment approach} that enhances \revision{smartphone mobile} readability by dynamically adjusting font parameters based on real-time contextual changes. Using smartphone sensors and a human-in-the-loop approach, SituFont personalizes text presentation to accommodate personal factors (e.g., fatigue, distraction) and environmental conditions (e.g., lighting, motion, location). To inform its design, we conducted formative interviews (N=15) to identify key SVI factors and controlled experiments (N=18) to quantify their impact on optimal text parameters. A comparative user study (N=12) across eight simulated SVI scenarios demonstrated SituFont’s effectiveness in improving \revision{smartphone mobile} readability \revision{in terms of improved efficiency and reduced workload compared with a non-trivial manual adjustment baseline}.

\end{abstract}


\begin{teaserfigure}
  \includegraphics[width=\textwidth]{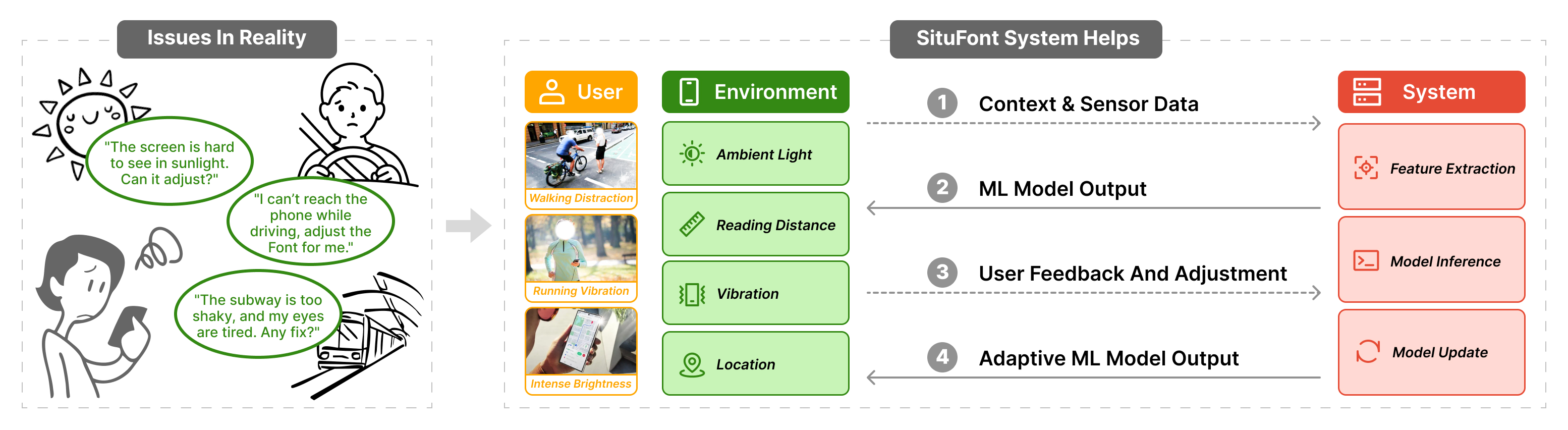}
  \caption{The SituFont system dynamically adapts font parameters based on user context and environmental conditions. The process involves four steps: (1) collecting context and sensor data, (2) generating ML model output, (3) incorporating user feedback for font parameter adjustments, and (4) producing adaptive model output for continuous improvement.}
  \Description{SituFont System Overview}
  \label{fig:teaser}
\end{teaserfigure}

\maketitle

\section{Introduction}
Mobile computing presents unique challenges due to its dynamic usage contexts \cite{sears2003when}. Unlike desktop computing, mobile devices are used in varied environments, leading to Situationally Induced Impairments and Disabilities (SIIDs) \cite{sears2003when, tigwell2019addressing}. Among these, Situational Visual Impairments (SVIs) arise from factors like low lighting and user motion, significantly affecting text readability \cite{sarsenbayeva2016situational, sarsenbayeva2017challenges, mustonen2004examining}. These challenges lead to visual fatigue, distraction, and difficulty processing information \cite{tigwell2018not}. 

Current solutions for addressing SVIs include manual adjustments \cite{hong2020implementation}, such as increasing font size \cite{huang2019effects}, using auditory substitutes \cite{carter2009seereader, hincapie2013crashalert, soubaras2020voice}, and automatic adjustments \cite{hong2020implementation}. 

While prior work has explored static solutions such as UI brightness modeling~\cite{evans2023measuring2}, post-hoc coping strategies for low visibility~\cite{tigwell2018not}, and the effects of font parameters on readability~\cite{huang2019effects, calvin2021image}, these approaches typically examine individual factors in isolation and rely on fixed or manually triggered adjustments~\cite{liu2024human}. However, real-world situational visual impairments (SVIs) often involve multiple dynamic contextual factors, such as motion, lighting variation, distraction, and fatigue, that interact and fluctuate over time. Under such conditions, static accessibility settings and simple rule-based interactions (e.g., manually increasing text size or pinch-and-zoom) place continuous attentional and interactional burden on users when visual conditions change rapidly or overlap~\cite{kong2025supporting}. Users need to repeatedly detect readability breakdowns and intervene manually, interrupting reading flow and increasing cognitive load.

This gap highlights the need for dynamic and personalized adaptation mechanisms that can respond to changing contexts and learn from user behavior over time. To address this need, we focus on smartphone reading in everyday dynamic environments, such as walking, commuting, or moving across diverse lighting and motion conditions~\cite{kong2025supporting, kong2023improving, zhang2024advancements, poslad2011ubiquitous}, a setting that is highly representative given the prevalence of mobile reading in daily life~\cite{shutaleva2023youth, kuzmicova2020mreading, wang2016read, shimray2015overview}. Within this context, we focus on Chinese text as a deliberate design probe for examining mobile readability under situational visual impairments, given its dense character-based structure and widespread use on smartphones, while recognizing that typographic adaptation may differ across alphabetic or bidirectional scripts~\cite{goldenberg2021towards}. Building on prior research in adaptive typography~\cite{kong2025supporting} and situational visual impairments~\cite{liu2024human, tigwell2018not}, we propose SituFont, an adaptive typography approach that leverages context-aware sensing and human-in-the-loop personalization to adjust font parameters just in time. SituFont leverages smartphone sensing to infer situational context and supports incremental personalization through user-initiated adjustments, which are incorporated as feedback to refine adaptation over time.

\revision{To ground the design of SituFont, we conducted two user studies to investigate the behavior patterns and SVI factors in smartphone mobile reading. We first conducted a semi-structured interview (N=15) to identify users' practices, challenges, coping mechanisms, and expectations in smartphone mobile reading. Results found three main categories (environmental, personal, and informational) of SVI factors and indicated typography adjustment as a general solution, but with personalized needs. We further conducted a controlled experiment (N=18) to examine how key environmental SVI factors influence the adjustment of key typography parameters. Results showed that different combinations of SVI factors led to desired preferences towards certain typography parameter trends.}

\revision{Building upon these study results, we grounded the system design and implementation of SituFont. First, we employed a label tree structure as an efficient representation of context, using hierarchical contextual labels that can be customized by the user. Based on such label sets, a machine learning model is adopted to predict the optimal font parameters based on the current context representation. Moreover, a human-in-the-loop interface is designed to collect individual adaptation feedback that can be progressively used to personalize the prediction model.} To evaluate SituFont’s effectiveness, we conducted a comparative study (N=12) simulating diverse SVI scenarios. Results show that SituFont improves reading efficiency, reduces perceived workload, and enhances users’ awareness of readability factors \revision{comparing with a non-trivial manual adjustment baseline where the user adjusts the parameters to a static optimal according to the scenario before the reading tasks.}

Building on prior work in adaptive typography and situational visual impairments, this paper makes the following contributions:

\begin{itemize}[topsep=0pt, parsep=0pt, partopsep=0pt]
\item Empirical insights from qualitative and quantitative studies examining how environmental, personal, and informational factors jointly affect mobile reading under SVIs.

\item The design and evaluation of SituFont, a population-informed, human-in-the-loop approach for just-in-time typographic adaptation that supports incremental personalization in dynamic contexts.

\item Design considerations for SVI interventions that emphasize the timing, activation, and integration of adaptation into the reading flow.
\end{itemize}

\section{Related Work}

\subsection{Situational Visual Impairments and Mitigation Strategies}

Situation-induced disorders and disabilities (SIIDs), introduced by Sears et al.~\cite{sears2003when}, describe contexts in which environmental, application-specific, or human factors impair user interaction. Sarsenbayeva et al. identified six key contributors to such barriers in mobile use: ambient temperature~\cite{sarsenbayeva2016situational,goncalves2017tapping}, ambient light~\cite{tigwell2019addressing}, environmental noise~\cite{barnard2007capturing}, mobility status~\cite{mizobuchi2005mobile,lin2007how,schildbach2010investigating,goel2012walktype}, burden~\cite{ng2013impact,ng2014investigating}, and pressure~\cite{lin2007how}. These factors disrupt interaction by degrading motor control and environmental awareness (e.g., movement, burden, temperature) or by hindering information acquisition (e.g., light, noise, attention). From a perceptual perspective, such situational factors reduce effective visual span and increase visual crowding, constraining reading rate and recognition under suboptimal viewing conditions~\cite{pelli2007crowding,rayner1998eye}. Among these, situational visual impairments (SVIs) are particularly impactful, directly reducing legibility and increasing visual strain on mobile displays~\cite{tigwell2019addressing,tigwell2018designing}. SVIs commonly arise from low lighting or motion, which impair reading comprehension and visual performance~\cite{mustonen2004examining,vadas2006reading,majrashi2022performance}. Ambient lighting~\cite{tigwell2018not,evans2023measuring} and environmental noise~\cite{sarsenbayeva2018effect,martin1988reading} further degrade text recognition and cognitive function.

To mitigate SVIs, prior work has explored compensatory strategies, particularly for “reading on the move,” where users must divide attention between navigation and reading. Auditory feedback has been proposed as a workaround~\cite{khan2020designing,yu2011enhancing,vadas2006reading,carter2009seereader}, but its linear structure and reliance on the auditory channel limit flexibility and usability~\cite{khan2020designing}. Other systems emphasize safety via visual augmentation; for example, CrashAlert~\cite{hincapie2013crashalert} uses depth cameras to detect obstacles. In more static contexts, design guidelines address SVIs under ambient light: Evans~\cite{evans2023measuring2} proposed a brightness perception model for UI design, and Tigwell~\cite{tigwell2018not} identified common SVI scenarios and coping strategies through a model linking environment, device, and user interaction. \revision{Complementary work on alternative display modes (e.g., light and dark modes) examines how global UI color schemes are adopted to improve comfort and accessibility across luminance conditions~\cite{andrew2024darkmode}.}
Despite these contributions, most approaches rely on static settings, predefined rules, or post-hoc compensation, offering limited real-time adaptation to dynamic environments. They also tend to overlook individual differences and user preferences in text perception, constraining personalization and broader applicability.

\subsection{Font Characteristics and Mobile Phone Text Legibility}

Optimizing text for SVIs requires understanding how font characteristics influence legibility on smartphones. Research highlights font size, weight, line spacing, and character spacing as key factors. Larger fonts improve readability, particularly for users with impaired vision, but excessively large sizes reduce efficiency by increasing scrolling demands \cite{huang2019effects,zhu2021effects,huang2009effects}. Bold fonts enhance recognition but may cause discomfort if too thick \cite{calvin2021image}. Proper line and character spacing improve layout clarity and character differentiation, making text easier to read \cite{zhu2021effects,wang2008chinese,ma2017interword,oralova2021spacing}. \revision{These typographic effects can be interpreted through visual-span and crowding accounts of reading, which show that letter size and spacing jointly constrain maximum reading rate under limited visual resolution and increased clutter \cite{pelli2007crowding,dobres2018crowding}.}

Beyond these font characteristics, additional factors such as screen size, text orientation, and reading direction also affect legibility. While larger screens generally facilitate easier reading, research suggests they do not significantly impact reading efficiency \cite{wang2013screen}. Similarly, text orientation and reading direction influence legibility, but their effects are often less pronounced than expected \cite{wigdor2005empirical,grossman2007exploring}. Dynamic text presentation methods, such as Rapid Serial Visual Presentation (RSVP) and peripheral vision considerations, have also been explored to enhance reading speed and efficiency \cite{maruya2012yubiyomu,uetsuki2017reading,hantani2022study}.
\revision{While prior research has explored how font characteristics affect readability, most studies focus on general legibility principles rather than their impact under situational constraints such as motion or lighting, and they are often conducted in controlled laboratories rather than realistic SVI contexts. As a result, it remains unclear how established typographic guidelines should be adjusted when readers must cope with motion, glare, or divided attention on mobile devices.}

\subsection{Just-in-time Adaptive Interventions with Human in the Loop}

Just-in-time adaptive interventions (JITAIs) provide personalized support by dynamically adapting to a user’s internal state and external context \cite{Nahum-Shani2018}, making them particularly relevant for mitigating SVIs where rapid environmental changes disrupt readability. While JITAIs have been widely applied in health domains \cite{Mair2022, Park2023, Hardeman2019, Pulantara2017, Goldstein2018, Terzimehic2017}, their use for addressing SVIs remains largely unexplored. JITAIs can be categorized as rule-based or AI-based: rule-based systems like FOCUS \cite{BenZeev2014} rely on predefined triggers and user input \cite{Gustafson2014}, while AI-driven JITAIs personalize interventions using behavioral and contextual data \cite{Time2Stop2024, Saponaro2021}. For example, Matthew et al. used a random forest model to predict nudge receptiveness from individualized context \cite{Saponaro2021}. Recent systems such as Time2Stop \cite{Time2Stop2024} incorporate human-in-the-loop feedback, refining strategies over time while maintaining transparency through AI-generated explanations.

\revision{Outside JITAIs, sensor-based adaptive mobile user interfaces have long explored how to personalize layout and interaction based on sensed context \cite{schmidt1999sensor,iqbal2018towards}, yet many of these systems have seen limited deployment, in part due to concerns around intrusiveness, transparency, and cost.} Advances in situational awareness technologies further extend the potential of JITAIs for SVIs. Mobile sensors can capture ambient light and motion data \cite{otebolaku2016user, calvin2021image, zhuang2019human}, while image recognition techniques extract semantic context from visual surroundings \cite{sharma2020image, forsstrom2014estimating, zheng2018weakly}. The integration of Large Language Models (LLMs), which excel at text understanding and generation \cite{vaswani2017attention}, enables context-aware interventions such as automatic font adjustments and personalized reading recommendations \cite{openai2024, zhu2023minigpt, liu2023improved}. \revision{At the same time, mobile sensing for adaptive interfaces raises well-documented privacy and acceptability concerns, prompting proposals for user-centered privacy controls and privacy-preserving processing of sensor data \cite{bemmann2023userprivacy,hernandez2020privacyca}.} Despite these advances, most JITAI systems focus on behavioral outcomes and overlook perceptual and cognitive challenges like SVIs. Furthermore, current AI-driven solutions often lack effective mechanisms for balancing automation and user control, raising concerns about intervention timing, transparency, and personalization.

\begin{figure*}[t]
  \centering
  \includegraphics[width=\textwidth]
  {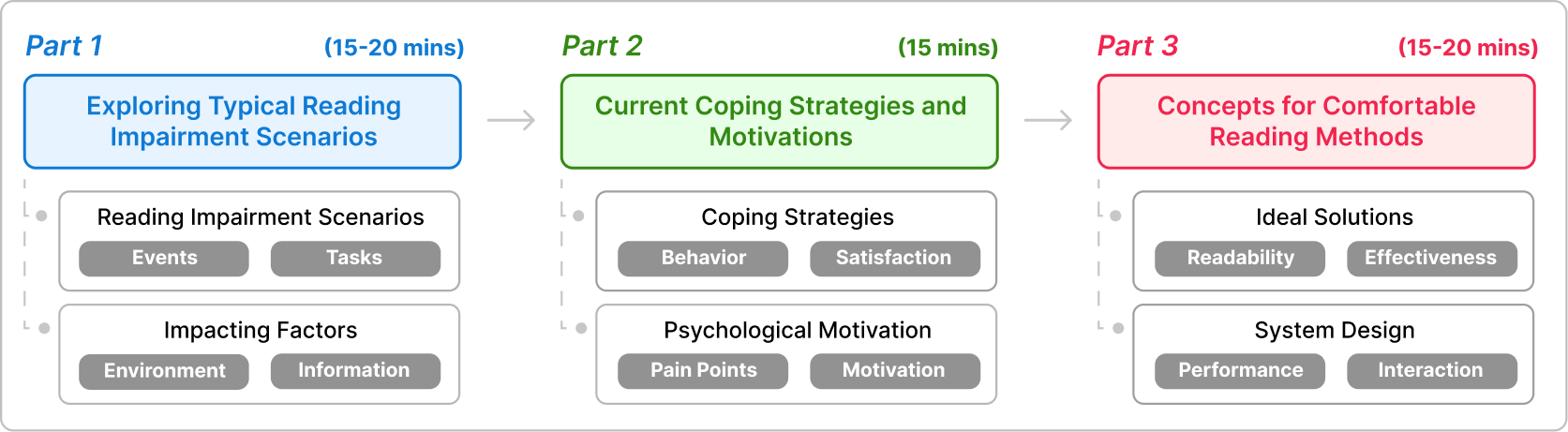}
    \caption{Workflow of the Semi-Structured Interviews.}
    \Description{Workflow of the Semi-Structured Interviews.}
    \label{fig:interview}
\end{figure*}

\begin{figure*}[t]
  \centering
  \includegraphics[width=\textwidth]{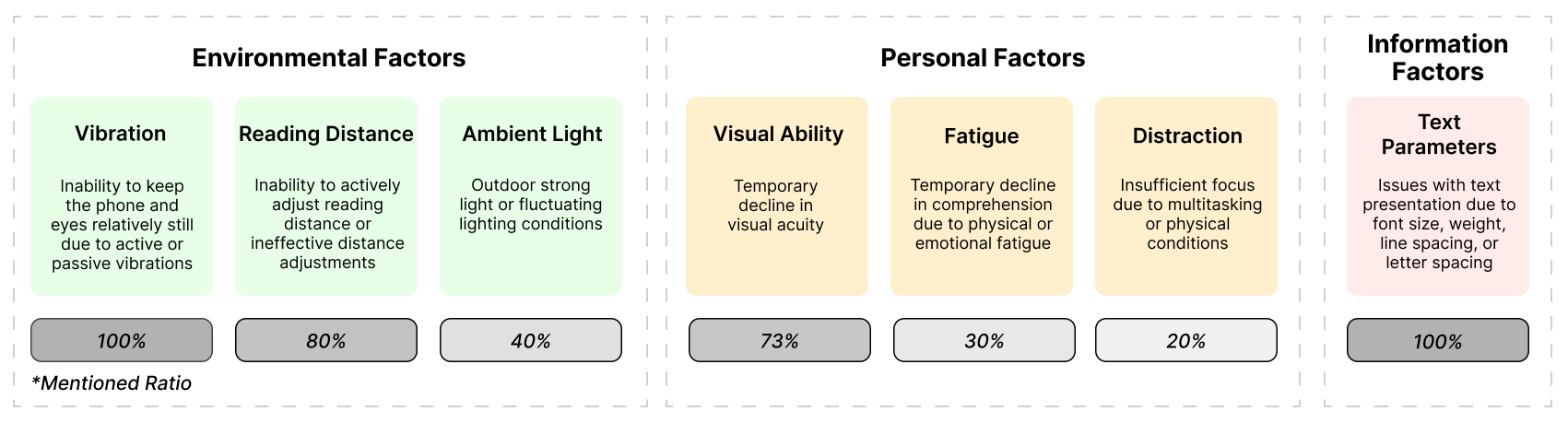}
    \caption{Interview Findings: Factors Affecting Situational Visual Impairments. The numbers inside the rectangles indicate the proportion of respondents who mentioned each factor.}
    \Description{Interview Findings: Factors Affecting Situational Visual Impairments. The numbers inside the rectangles indicate the proportion of respondents who mentioned each factor.}
    \label{fig:SVI_factors}
\end{figure*}

\section{Formative study}
\label{sec:formative_study}

\revision{To design a system that mitigates SVIs in mobile reading, we operationalize the “Understanding – Sensing – Modeling – Adapting” cycle for situational visual impairments and related SIIDs in smartphone reading \cite{tigwell2019addressing}. In this framework, the \textit{Understanding} module identifies contextual factors that affect readability, which are then detected by the \textit{Sensing} module. The \textit{Modeling} module determines how these factors influence reading, enabling the \textit{Adapting} module to dynamically adjust text presentation. In the following sections, we first contextualize key SVI-inducing factors and users' existing coping strategies for Chinese mobile reading through Study 1 (Section~\ref{section: Study 1}), then examine how to model these factors and the associated font adjustments for real-time adaptation in Study 2 (Section~\ref{subsection: Study 2}), and finally synthesize the findings into design implications that inform the sensing, modeling, and interaction mechanisms of SituFont.}

\subsection{Study 1: Contexts, Factors, and Coping Strategies}
To identify environmental and personal factors that induce SVIs during mobile reading, as well as user strategies for mitigation, we conducted 15 semi-structured interviews to explore participants' challenges, coping mechanisms, and 
\revision{expectations for more supportive reading interfaces in everyday situations.}

\label{section: Study 1}

\subsubsection{Research Methods}
We interviewed 15 participants aged 19–52, including students and professionals with diverse visual conditions such as myopia, astigmatism, presbyopia, and normal vision (Appendix \ref{appendix: Study 1}, Table \ref{tab:participants}). Participants were recruited via university mailing lists and public social media posts. Interviews were conducted remotely using Tencent Meeting\footnote{\url{https://meeting.tencent.com/}} and lasted 30–40 minutes. Each session began with an overview of SVIs, followed by discussions of participants’ mobile reading experiences, coping strategies, and suggestions for improving readability (Figure \ref{fig:interview}; Appendix \ref{appendix: Study 1}). 
All interviews were audio-recorded, transcribed, and analyzed in MAXQDA \cite{maxqda2022} using inductive thematic analysis \cite{braun2006using}. Two researchers independently conducted open coding, iteratively refined a shared codebook, and clustered conceptually related codes to generate final themes. \revision{Disagreements were resolved through discussion until consensus was reached.}

\subsubsection{Findings}

\textit{Environmental Factors.} Reading while in motion posed a major challenge, especially during walking (11/15) and running (9/15). Movement-induced vibration blurred text, impairing recognition (P3, P6, P8–9, P13, P15), and sometimes caused dizziness (P8, P10, P13) or slower reading speeds (P6, P12, P14–15). Some participants (P4) found it difficult to concentrate due to the need to monitor their surroundings, while others (P1, P5, P11) avoided reading while running altogether. In addition, all participants (15/15) reported difficulty reading in vehicles, especially during commutes (P6, P7, P10). Movement in private cars disrupted focus and sometimes led to motion sickness (P2, P7, P10, P12). Bright outdoor light was another common issue. Six participants (P1–2, P10–11, P13–14) noted that screen brightness adjustments were often insufficient under direct sunlight, necessitating increased font size and weight to maintain legibility. Reading at suboptimal viewing distances was also problematic, particularly when users could not freely adjust their phone position. This occurred in scenarios such as driving, where phones were mounted at a fixed distance (P5), or crowded environments like subways (P13), where users lacked space to reposition their device. \revision{These four primary SVI factors, which include intense brightness, high vibration, distraction, and fatigue, have also been widely reported in prior work on SVIs and mobile readability. Rather than proposing new SVI categories, our contribution lies in characterizing how these factors co-occur in everyday mobile reading, and in operationalizing their most realistic combinations into design-relevant scenarios for adaptive typography and JITAI-based interventions.}

\textit{Personal Factors.} Visual acuity impacted readability, especially for participants who wore glasses or contact lenses (9/11). Challenges were exacerbated when they were not wearing corrective lenses, particularly before bed (P4, P6, P7, P10, P11, P12). Comprehension ability influenced ease of reading, with some participants (P8, P14) reporting increased visual fatigue when reading lengthy texts. Concentration difficulties were another barrier, as participants (P3, P10, P13) mentioned distractions and multitasking negatively affected their reading efficiency. \revision{These personal factors later informed the personalized branch of our context label hierarchy and the design of user-configurable preferences in SituFont.}

\textit{Information Factors.} Participants generally agree that text presentation played a crucial role in readability. Text parameters such as excessively small font sizes ($n = 13$), thin fonts ($n = 6$), overly large line spacing ($n = 4$), and small character spacing ($n = 2$) were frequently cited as readability obstacles. Information difficulty was another key factor as highly specialized or information-dense content was difficult to comprehend ($n = 2$). \revision{These observations motivated our focus on font parameters as the primary adaptation channel, while treating content difficulty as a separate factor that is more appropriately addressed through content selection or summarization rather than low-level typography alone.}

\textit{Strategies and Challenges.} Adjusting reading distance was the most commonly reported strategy, though it often caused physical discomfort. For instance, P6 described repeatedly moving their phones closer or farther, which led to neck strain during prolonged use. To reduce visual load, several participants (P8, P10, P13, P15) used auditory substitutes like voice assistants or text-to-speech tools, particularly when visual input was compromised. However, these were frequently impractical. P13 noted inefficiency while driving, and others (P9, P15) cited issues in noisy or private environments. All participants (15/15) increased font size as a workaround, but many, especially younger users (8/15), found that it disrupted formatting and required excessive scrolling. There was also widespread reluctance to manually adjust settings due to the time and effort involved; participants expressed a preference for systems that could automatically adapt to changing conditions without user input. These findings highlight the demand for seamless, context-aware font adjustment 
\revision{that complements, rather than replaces, users' existing strategies, and they directly motivate our human-in-the-loop approach in later system design.}

\subsection{Study 2: Modeling Key Factors for Adaptation} \label{subsection: Study 2}

Building on qualitative insights from Study 1, we identified three major categories of factors influencing SVIs: environmental conditions (e.g., motion, lighting), personal characteristics (e.g., visual acuity, comprehension, concentration), and information-related aspects (e.g., text size, spacing, content complexity, length) (Figure \ref{fig:SVI_factors}). These findings informed the design of Study 2, which aimed to quantify how environmental conditions affect 
\revision{preferred font parameters and to provide aggregate data that can be used as an initial model for JITAI-based systems before sufficient individual data become available.} We conducted a controlled experiment examining how light intensity, motion states, and vibration levels impact adjustments to key 
\revision{font} parameters, including font size, weight, line spacing, and character spacing. 

\subsubsection{Research Methods}

\textit{Participants.} We recruited 18 university students aged 18 to 25 with normal or corrected vision through public social media posts; none had participated in the prior interviews. \revision{Consistent with the scope of this work, participants reported daily experience reading Chinese text on smartphones.} Each participant experienced six scenarios combining two light intensity levels (low, high) and three movement states (standing, walking, running) (Table \ref{tab: Experimental Scenarios}). 
\revision{The indoor standing condition served as a low-SVI baseline, and the remaining five conditions combined increased motion and/or outdoor light to approximate common real-world SVI situations (e.g., commuting, walking outdoors).}

\begin{table*}[h!]
  \begin{tabular}{ccccc}
    \toprule
    ID & Scenario & Motion State & Light Level & Vibration Level \\
    \midrule
    1 & Indoor Corridor & Standing & Low & Low \\
    2 & Indoor Corridor & Walking & Low & Medium \\
    3 & Indoor Corridor & Running & Low & High \\
    4 & Outdoor Playground & Standing & High & Low \\
    5 & Outdoor Playground & Walking & High & Medium \\
    6 & Outdoor Playground & Running & High & High \\
    \bottomrule
  \end{tabular}
\caption{Experimental Scenarios: Motion, Light, and Vibration Conditions}
  \label{tab: Experimental Scenarios}
\end{table*}

\textit{Procedure. }Participants used a custom-built mobile application preloaded with six isomorphic Chinese reading passages (550 characters each, high school level) and displayed in plain text. They read the passages in both indoor and outdoor environments while standing, walking, or running. Participants were instructed to stand still, walk at a natural pace, and run at a light jogging pace. \revision{Before the experiment, we instructed participants to prioritize legibility and comfort rather than speed, and to adjust font parameters until they felt the text was comfortably readable in each scenario.} During reading, participants adjusted 
font parameters (font size, weight, line spacing, and character spacing) via a one-handed interface. Adjustments could be made freely at any time by double-tapping the screen. During the experiment, indoor lighting was strictly controlled to be below 200 lux (with an average of approximately 100 lux), while outdoor lighting was strictly maintained above 10,000 lux (with an average of approximately 30,000 lux).

\textit{Apparatus. }To ensure consistency, we used two HUAWEI P40 smartphones (6.1-inch screen, 2340×1080 resolution) across all sessions. Built-in sensors recorded ambient light, acceleration (vibration), and reading distance. Display settings were standardized: black text on a white background, with brightness fixed at 100\% for outdoor (corresponding to 531 nits) and 50\% for indoor conditions. These settings were uniformly applied to both devices. Environmental data (light intensity, reading distance, vibration offset) were recorded alongside participants’ 
\revision{font} adjustments. Line and character spacing were logged in \textit{em} units (e.g., 0.05\textit{em} = 5\% of font size). Data cleaning involved removing incomplete or clearly erroneous entries, such as missing sensor data or implausible reading distances caused by blur or sensor errors. Thresholds for exclusion were empirically defined based on pilot testing. After cleaning, 497 valid datasets remained, averaging 28 per participant and 83 per scenario. \revision{Study 2 did not evaluate SituFont directly; instead, its role was to characterize how participants adjust font parameters under controlled manipulations of light and motion, and to provide design priors for later modeling and adaptation.} Additional technical and procedural details are provided in Appendix \ref{appendix:Experiment}.

\subsubsection{Findings}

\begin{table*}[t]
\centering
\begin{threeparttable}
\small 
\begin{tabularx}{\textwidth}{cXcccccc}
\toprule
ID & Scenario & \textit{X}-Axis (\textit{m/s\textsuperscript{2}}) & \textit{Y}-Axis (\textit{m/s\textsuperscript{2}}) & \textit{Z}-Axis (\textit{m/s\textsuperscript{2}}) & Light (\textit{lux}) & Reading Distance (\textit{cm}) \\
\midrule
1 & Indoor Standing & $0.18 \pm 0.13$ & $0.11 \pm 0.08$ & $0.26 \pm 0.22$ & $111.69 \pm 89.14$ & $33.00 \pm 11.46$ \\
2 & Indoor Walking & $0.53 \pm 0.15$ & $0.65 \pm 0.15$ & $0.80 \pm 0.28$ & $86.08 \pm 81.86$ & $31.84 \pm 9.69$ \\
3 & Indoor Running & $1.41 \pm 0.59$ & $1.42 \pm 0.85$ & $1.58 \pm 0.55$ & $82.42 \pm 62.78$ & $30.99 \pm 8.73$ \\
4 & Outdoor Standing & $0.16 \pm 0.08$ & $0.10 \pm 0.05$ & $0.24 \pm 0.13$ & $51208.67 \pm 27990.79$ & $29.96 \pm 10.98$ \\
5 & Outdoor Walking & $0.52 \pm 0.14$ & $0.68 \pm 0.14$ & $0.78 \pm 0.19$ & $38263.15 \pm 23249.25$ & $30.84 \pm 9.03$ \\
6 & Outdoor Running & $1.54 \pm 0.55$ & $1.52 \pm 0.90$ & $1.70 \pm 0.55$ & $36482.24 \pm 25716.72$ & $29.33 \pm 8.57$ \\
\midrule
\textit{F} Value & & 261.146 & 121.825 & 258.395 & 125.669 & 1.496 \\
\textit{p} Value & & 0.000** & 0.000** & 0.000** & 0.000** & 0.189 \\
\bottomrule
\end{tabularx}
\begin{tablenotes}
\item * \textit{p} < 0.05 , ** \textit{p} < 0.01
\end{tablenotes}
\caption{Variance Analysis Table of Sensor Data Features in 6 Experimental Scenarios}
\label{tab:variance analysis}
\end{threeparttable}
\end{table*}

\begin{table*}[t]
\centering
\begin{threeparttable}
\begin{tabular}{lccccc}
\toprule
& Reading Distance & Light Intensity & \multicolumn{3}{c}{Vibration Offset} \\
\cmidrule(lr){4-6}
& & & \textit{X} Axis & \textit{Y} Axis & \textit{Z} Axis \\
\midrule
Font Size & 0.404** & 0.117** & 0.368** & 0.333** & 0.381** \\
Font Weight & 0.228** & 0.373** & 0.230** & 0.263** & 0.225** \\
Line Spacing & 0.132** & -0.038 & 0.170** & 0.168** & 0.173** \\
Character Spacing & 0.057 & 0.042 & -0.036 & -0.066 & -0.037 \\
\bottomrule
\end{tabular}
\begin{tablenotes}
\item * \textit{p} < 0.05 , ** \textit{p} < 0.01
\end{tablenotes}
\caption{Spearman Correlation Coefficients of Environmental Factors and Readable Text Parameters}
\label{tab:spearman}
\end{threeparttable}
\end{table*}

Significant differences were observed in light intensity and motion states across scenarios while reading distance remained stable (Table \ref{tab:variance analysis}). Light intensity was substantially higher in outdoor ($M = 41998.37, SD = 26421.78$) compared to indoor environments ($M = 93.90, SD  = 79.66$). Motion states also exhibited significant differences, confirmed by three-axis acceleration data. Standing showed the lowest values ($X: M=0.17, SD=0.10; Y: M=0.10, SD=0.07; Z: M=0.25, SD=0.18$), followed by walking ($X: M=0.53, SD=0.14; Y: M=0.67, SD=0.14; Z: M=0.79, SD=0.24$) and running ($X: M=1.48, SD=0.57; Y: M=1.48, SD=0.87; Z: M=1.65, SD=0.55$). Reading distance remained consistent across scenarios.

\begin{table*}[t]
\centering
\begin{threeparttable}
\begin{tabular}{lccccc}
\toprule
ID & Scenario & Font Size (\textit{sp}) & Font Weight (\textit{px}) & Line Spacing (\textit{em}) & Character Spacing (\textit{em}) \\
\midrule
1 & Indoor Standing & 20.60±3.99 & 0.59±0.65 & 0.25±0.14 & 0.10±0.10 \\
2 & Indoor Walking & 21.29±3.98 & 0.79±0.76 & 0.27±0.17 & 0.08±0.09 \\
3 & Indoor Running & 23.84±5.31 & 0.93±0.80 & 0.29±0.16 & 0.10±0.09 \\
4 & Outdoor Standing & 19.28±3.63 & 0.93±0.79 & 0.25±0.16 & 0.09±0.08 \\
5 & Outdoor Walking & 21.08±3.26 & 1.04±0.95 & 0.22±0.13 & 0.09±0.09 \\
6 & Outdoor Running & 23.64±4.92 & 1.30±0.87 & 0.31±0.14 & 0.12±0.14 \\
\midrule
\textit{F} Value & & 15.108 & 7.275 & 4.666 & 1.301 \\
\textit{p} Value & & 0.000** & 0.000** & 0.000** & 0.262 \\
\bottomrule
\end{tabular}
\begin{tablenotes}
\item * \textit{p} < 0.05 , ** \textit{p} < 0.01
\end{tablenotes}
\caption{ANOVA Table of Readable Text Parameters in 6 Experimental Scenarios}
\label{tab:ANOVA}
\end{threeparttable}
\end{table*}

\begin{table*}[t]
\centering
\setlength{\tabcolsep}{4pt} 
\begin{tabular}{lccccc}
\toprule
ID & Scenario & Size (sp) & Weight (px) & Line (em) & Char (em) \\
\midrule
1 & Indoor Standing & 2.70 & 0.65 & 0.14 & 0.09 \\
2 & Indoor Walking  & 2.94 & 0.66 & 0.16 & 0.08 \\
3 & Indoor Running  & 4.47 & 0.69 & 0.16 & 0.09 \\
4 & Outdoor Standing & 2.08 & 0.61 & 0.13 & 0.06 \\
5 & Outdoor Walking & 2.40 & 0.83 & 0.13 & 0.08 \\
6 & Outdoor Running & 3.80 & 0.79 & 0.12 & 0.13 \\
\bottomrule
\end{tabular}
\caption{Standard deviation of preferred typographic parameters across six experimental scenarios (N=18).}
\label{tab:std-params}
\end{table*}

\textit{Light Intensity: Low v\st{.}s. High.}
Our analysis revealed that light intensity significantly influenced font weight. In low-light indoor conditions, participants maintained relatively consistent font weights ($M = 0.59$ indoor standing, $M = 0.79$ indoor walking).  In contrast, in high-light conditions (outdoor), participants consistently increased font weight to enhance text contrast, with the strongest adjustments seen in outdoor running ($M = 1.30, SD = 0.87$). Font weight showed a strong positive correlation with light intensity ($r = 0.373, p < 0.01$). Font size showed relatively minor adjustments under varying light conditions ($r = 0.117, p < 0.01$), indicating that participants primarily relied on weight adjustments rather than size to maintain readability.

\textit{Motion: Low v\st{.}s. High.}
Motion had a pronounced impact, particularly on font size. In low-motion scenarios, participants used smaller font sizes ($M = 20.60, SD = 3.99$ indoor). In high-motion scenarios, participants significantly increased font size ($M = 23.84, SD = 5.31$ indoor, $M = 23.64, SD = 4.92$ outdoor). Font size showed a strong positive correlation with motion ($r = 0.368, p < 0.01$), reflecting the need for larger text to maintain clarity during movement.Line spacing also exhibited moderate adjustments in high-motion scenarios, showing significant positive correlations with vibration ($r = 0.170, p < 0.01$). However, character spacing remained largely unaffected.

\textit{Vibration: Low v\st{.}s. High.}
Vibration particularly impacts font size and font weight. In low-vibration scenarios (standing), participants used smaller font sizes ($M = 20.60, SD = 3.99$ indoor). However, in high-vibration scenarios (running), font size ($M = 23.84, SD = 5.31$ indoor, $M = 23.64, SD = 4.92$ outdoor) and font weight ($M = 1.04, SD = 0.95$ indoor, $M = 1.30, SD = 0.87$ outdoor) increased significantly. Vibration showed strong correlations with font size ($r = 0.360, p < 0.01$) and font weight ($r = 0.239, p < 0.01$).

\textit{High-Light + Motion.}
In scenarios combining high light and motion, participants made the most significant adjustments in both font size and weight. For instance, in outdoor running scenarios, font size reached $M = 23.64, SD = 4.92$, and font weight increased to $M = 1.30, SD = 0.87$ (Table \ref{tab:ANOVA}). \revision{These patterns suggest that high luminance and motion jointly motivate participants to increase both the size and contrast of text to preserve legibility.}

\textit{High-Light + Vibration.}
Under conditions combining high light and vibration, font weight was the most adjusted parameter ($r = 0.373, p < 0.01$), emphasizing contrast as the critical factor.

\textit{Individual Differences.}
Text adjustments varied significantly across participants, reflecting individual preferences (Table \ref{tab:std-params}). Font size variability was highest in indoor running scenarios ($SD = 4.467$). Font weight variability peaked in outdoor walking ($SD = 0.827$) and outdoor running ($SD = 0.791$). Line spacing showed moderate variability, while character spacing exhibited minimal changes. \revision{These variations motivated us to treat the relationships between environmental conditions and font parameters derived from Study 2 as population-level priors that can be further personalized in later adaptive models, rather than as prescriptions for a single optimal configuration.}

\subsection{Design Implications}
\label{section: Design Implications}

\revision{Findings from Study 1 (interviews) and Study 2 (controlled experiment) converge on several design implications for readability support under SVIs. These implications informed the sensing, modeling, and interaction mechanisms of SituFont.}

\textit{Contextual Awareness and Automated Adaptation.} Both studies highlight the importance of context-aware font adjustments \revision{grounded in sensed environmental conditions}. In Study 1, participants 
\revision{participants expressed a preference for systems that adjust text presentation in response to changing conditions, reducing the need for repeated manual configuration.}
\revision{Study 2 quantified how motion and lighting are associated with systematic changes in preferred font size and weight, particularly} in high-motion (e.g., running) and high-light (e.g., outdoor) scenarios. 
\revision{Together, these findings suggest that real-time sensing via smartphone sensors (e.g., accelerometers, ambient light) can provide a basis for automated adjustments that approximate users' preferred configurations under different SVIs.}

\textit{Personalization and User Agency.} Study 1 emphasized the need for customizable settings, with participants noting that "\textit{everyone’s eyes are different}." Study 2 confirmed this through large individual variations in font size and weight preferences, especially under demanding conditions. 
\revision{These observations indicate that population-level models derived from aggregate data are insufficient on their own. Systems should therefore offer multiple adaptation levels, such as adjustable sensitivity, optional prompts before changes, and the ability to override or refine suggestions, so that automated adaptations can be progressively personalized. A hybrid model that balances automation and user agency is likely to better accommodate diverse visual needs and preferences.}

\textit{Addressing Limitations of Existing Coping Strategies.} Study 1 participants reported using manual adjustments, larger fonts, and text-to-speech, but found them insufficient. Reading distance adjustments caused discomfort, large fonts disrupted layout, and auditory tools were impractical in noisy or private settings. Study 2 confirmed that no single adjustment resolved all challenges, highlighting the need for multifactorial adaptation strategies. \revision{These findings point toward combining adjustments to multiple font parameters (e.g., size, weight, spacing) in response to specific SVI patterns, rather than relying solely on single-parameter changes such as global zoom or font size increase.}

\textit{Simplified Interaction and Reduced Cognitive Load.} Participants in Study 1 found manual adjustments burdensome and often avoided them. To ease interaction, systems should 
\revision{minimize the need for deep navigation into settings menus and instead offer lightweight, context-aware interaction mechanisms. For example, gesture-based shortcuts (e.g., a long-press that triggers an adaptive suggestion) can provide quick access to adjustments, while adaptive interfaces can learn from users' confirmations and corrections over time, thereby} reducing repeated manual configuration.

\textit{Targeted Adaptations for High-Variability Scenarios.} Both studies point to the importance of tailoring interventions to high-variability scenarios, especially reading in motion, adapting to light changes, and accounting for user differences. Study 1 identified these as most disruptive; Study 2 confirmed that running and outdoor settings showed the greatest need for font adjustments. \revision{These patterns motivated us to treat such scenarios (e.g., outdoor running, outdoor walking) as distinct context nodes when structuring the label hierarchy and to prioritize them as primary targets for adaptive interventions in SituFont.}

\section{SituFont System Design and Implementation}

\begin{figure*}[t]
  \centering
  \includegraphics[width=\textwidth]{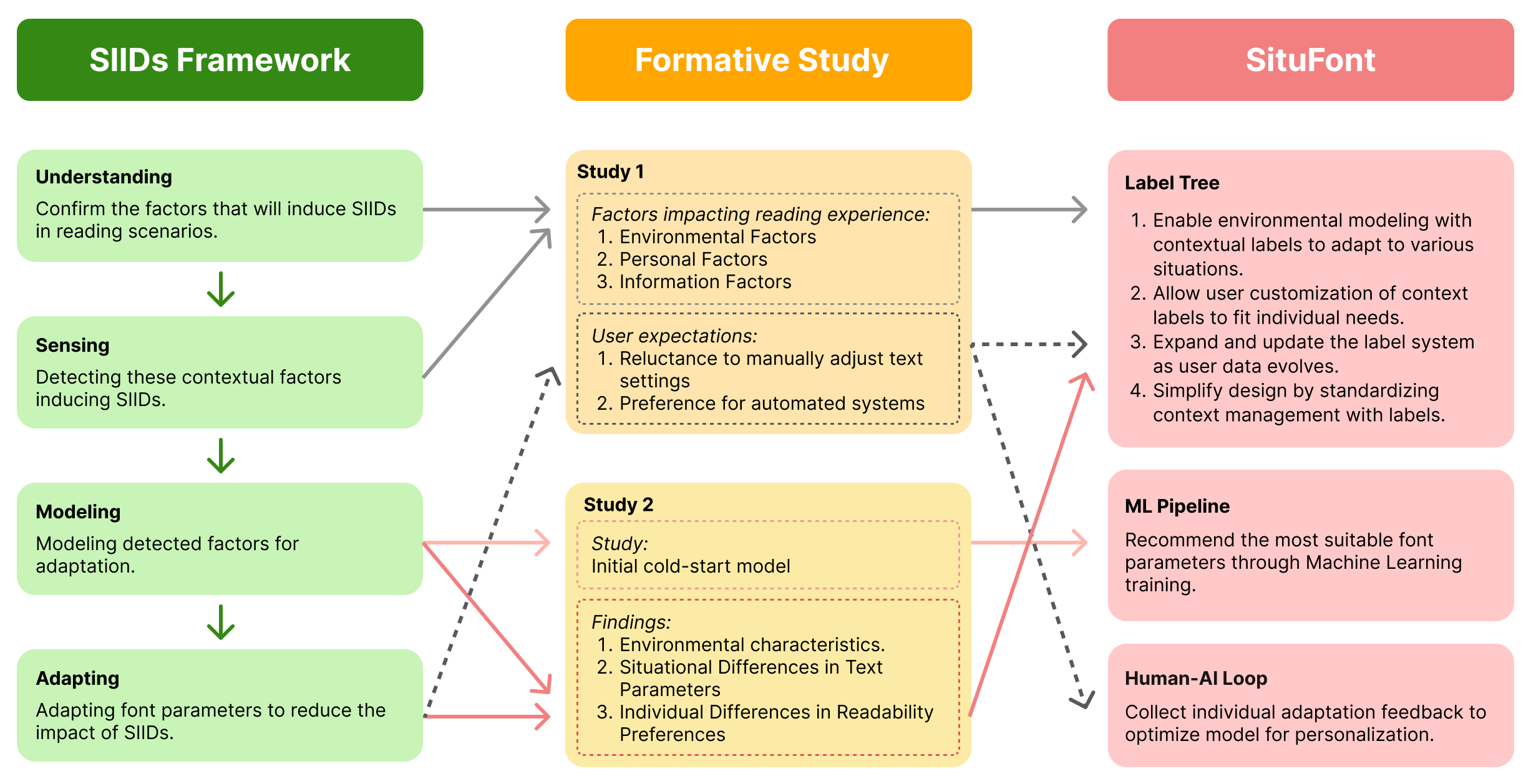}
    \caption{Formative study 's findings inspire the system design of SituFont, which mainly include the label tree, machine learning training, human-ai loop modules.}
    \Description{Formative study 's findings inspire the system design of SituFont, which mainly include the label tree, machine learning training, human-ai loop modules.}
    \label{fig:system_implications}
\end{figure*}

As \revision{summarised in} Figure \ref{fig:system_implications}  
\revision{, we used these formative insights to design SituFont, a JITAI-inspired system that enhances readability by dynamically adjusting font parameters based on real-time contextual changes and user feedback.} In the sections below, we outline the design (Section \ref{subsec:SituFont_design}) and implementation (Section \ref{subsec:SituFont_implementation}) of the SituFont system.

\subsection{SituFont System Design Overview} 
\label{subsec:SituFont_design}

SituFont's core components include the ML Training Pipeline, the Label Tree of Reading Scenario, and the Human-AI loop Workflow. Each module plays a key role in the system’s font adaptation and effectiveness.

\subsubsection{Machine Learning for Font Parameters Recommendation}
\label{subsubsec:Design_ML}
Constructing an ML-driven JITAI system for recommending suitable font parameters involves two steps:

\textit{(1) Initial Model from Group Data.}

SituFont initializes font parameter recommendations using an ML model trained on aggregated group-level adjustment data collected in Formative Study 2 (Section~\ref{subsection: Study 2}). This model captures common adaptation patterns observed across users as they adjusted font size, weight, letter spacing, and line spacing under varying environmental conditions such as reading distance, light intensity, and phone acceleration. The resulting predictions reflect population-level trends in how typography is adapted under SVIs, providing a reasonable cold-start behavior when users first interact with the system.

Developing this initial model serves two purposes. First, it enables the system to provide immediate, context-sensitive font recommendations without requiring prior personalization. Second, it establishes a foundation that can be incrementally refined with user-specific feedback over time, allowing the model to adapt to individual reading habits with minimal additional data. In this setup, environmental factors are used as input features, while adjusted typographic parameters serve as output variables for supervised learning.

\begin{table*}[t]
  \begin{tabular}{ccccc}
    \toprule
    Data Name & Unit & Component and Method & Frequency \\
    \midrule
    Ambient Light & \textit{lux} & Light Sensor; Direct Call & 10 times/sec \\
    Reading Distance & \textit{cm} & Front Camera; Calculated from pupil distance & 10 times/sec \\
    Vibration Offset & \textit{m/s\textsuperscript{2}} & Accelerometer; sensorEvent.values[0/1/2] for \textit{x/y/z} axis & 10 times/sec \\
    Font Size & \textit{sp} & Android function for size after each adjustment & Recorded on upload \\
    Font Weight & \textit{px} & Android function for weight after each adjustment & Recorded on upload \\
    Line Spacing & \textit{em} & Android function for spacing after each adjustment & Recorded on upload \\
    Letter Spacing & \textit{em} & Android function for spacing after each adjustment & Recorded on upload \\
    \bottomrule
  \end{tabular}
  \caption{Environmental Sensor Data collected in SituFont}
  \label{tab:environmental sensor}
\end{table*}

\begin{figure*}[t]
  \centering
  \includegraphics[width=\textwidth]{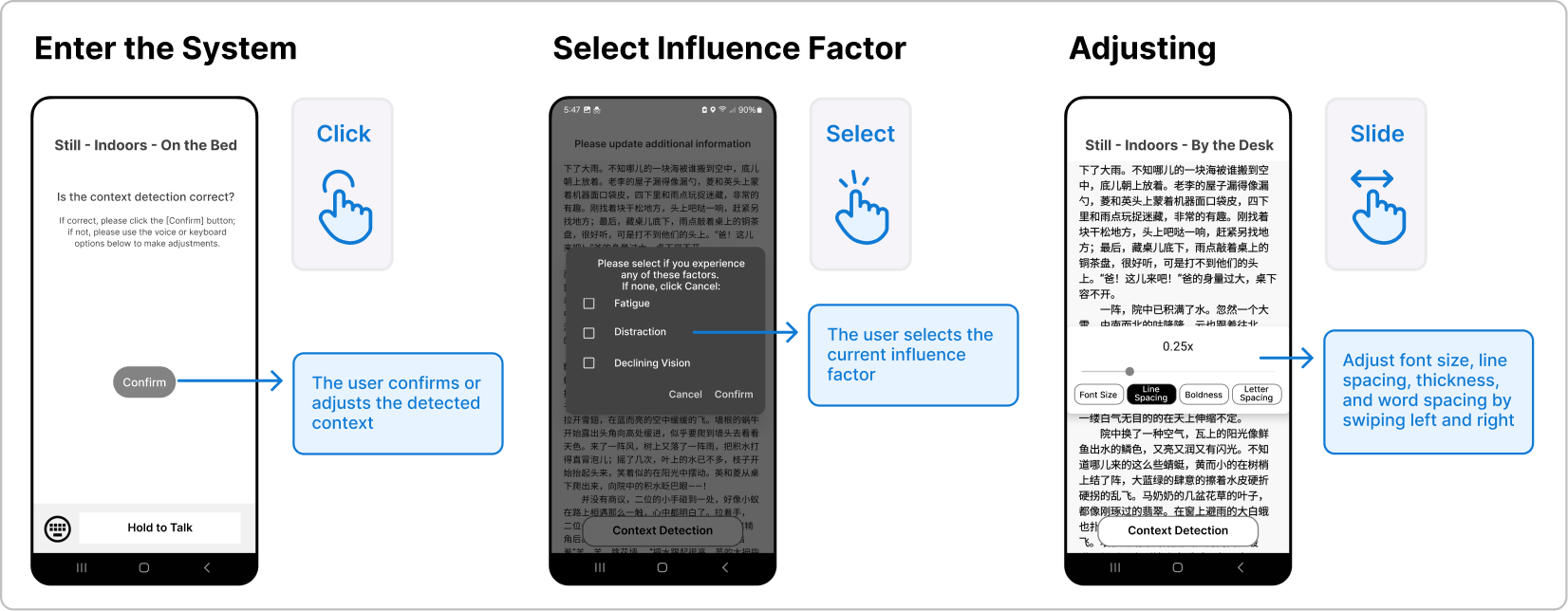}
    \caption{The user interface of SituFont involves three key interactions: (1) Entering the System – the user confirms or adjusts the detected reading context; (2) Selecting Influence Factors – the user specifies factors affecting readability, such as fatigue or distraction; and (3) Adjusting Text Parameters – the user refines font size, line spacing, thickness, and word spacing through swipe gestures for a personalized reading experience.}
    \Description{The user interface of SituFont involves three key interactions: (1) Entering the System – the user confirms or adjusts the detected reading context; (2) Selecting Influence Factors – the user specifies factors affecting readability, such as fatigue or distraction; and (3) Adjusting Text Parameters – the user refines font size, line spacing, thickness, and word spacing through swipe gestures for a personalized reading experience.}
    \label{fig:User interface}
\end{figure*}

\textit{(2) Collecting Data from Users’ Daily Usage.}
Data is collected through the interface shown in the system (figure \ref{fig:User interface}). When users double-tap the screen, a control panel for adjusting text parameters appears near the tapped location to make it easier to adjust settings. While users modify parameters, the system automatically detects reading distance, light intensity, and phone acceleration (Table \ref{tab:environmental sensor}). Users' current cognitive factors are collected by prompting them to choose whether any factors related to fatigue, distraction, and temporary decrease in visual ability exist. Once users click on a blank area of the screen, the data is sent to the backend to be stored in the corresponding contextual dataset, which is used to train and update the model for future recommendations.

\subsubsection{Label Tree of Reading Scenario}
\label{subsubsec:Label_Tree_Design}

\begin{figure*}[t]
  \centering
  \includegraphics[width=\textwidth]{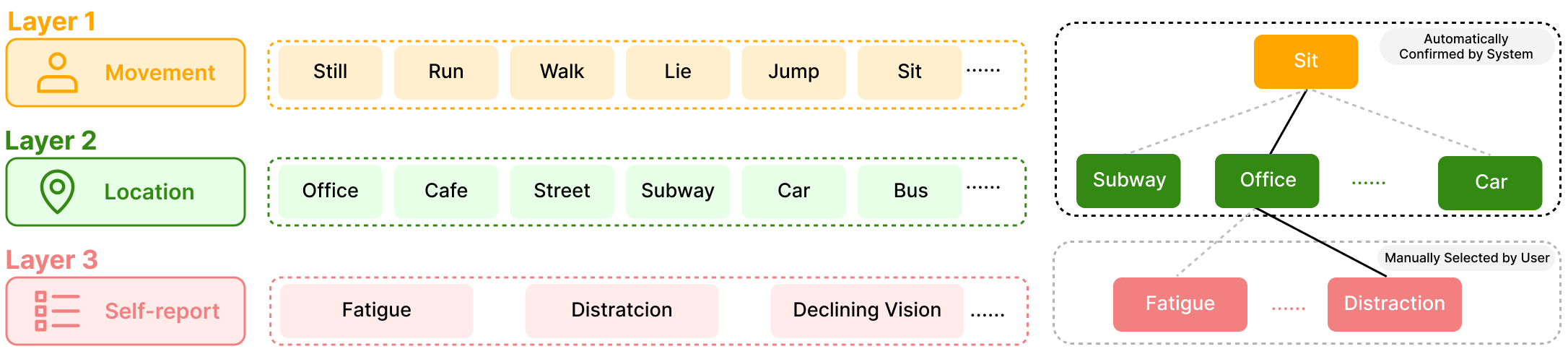}
    \caption{The left part of the figure describes a three-layer labeling system used to mark situations, where the priority decreases from top to bottom when constructing the label tree. The right part of the figure presents an example of a label tree, where the solid lines indicate the process of detecting situational label combinations.}
    \Description{The left part of the figure describes a three-layer labeling system used to mark situations, where the priority decreases from top to bottom when constructing the label tree. The right part of the figure presents an example of a label tree, where the solid lines indicate the process of detecting situational label combinations.}
    \label{fig:labeltree}
\end{figure*}

\textit{Label the tree structure.}
Based on findings from the formative study, we designed a hierarchical data structure for reading context labels (figure \ref{fig:labeltree}), categorized as "[Movement/Posture] - [Environmental Scene] - [Personalized Needs]." The system automatically determines the first two layers of context labels by sensing the user's current movement state, environment, and location. However, personalized factors (such as visual ability, fatigue, and attention state) that represent the user's individual conditions cannot be directly detected by the system. For example, "Has the user's vision changed?", "Is the user feeling fatigued?", and "Is the user focused?" are factors that are difficult for the system to assess automatically. Therefore, the third layer of labels requires the user to manually indicate whether their vision, comprehension, or attention is currently affected.

The hierarchy follows the order of [Movement/Posture] - [Environmental Scene] - [Personalized Needs] because the first two categories can be automatically detected by the system. Among these, Movement/Posture is considered more influential than Environmental Scene on reading behavior, so it is placed at the top of the hierarchy. However, personalized needs require manual input from the user and should be avoided as much as possible, which is why it is placed as the final layer in the label tree.

\textit{Label Tree Functions.}  
The Label Tree is designed to maximize the utility of small datasets by structuring them hierarchically based on contextual factors such as lighting conditions, user states, and task demands. This organization ensures that each dataset retains its relevance within its specific environment, allowing fine-tuned models to be applied effectively without requiring extensive data collection. Moreover, the Label Tree facilitates context transfer and generalization, enabling the system to identify the most relevant existing dataset when encountering a new but similar context. This reduces redundancy and enhances adaptability, allowing small datasets to be leveraged efficiently across multiple scenarios. By systematically preserving contextual distinctions and enabling knowledge transfer, the Label Tree significantly improves the effectiveness of small datasets, making them more impactful while minimizing the need for extensive user input or additional data collection.

\subsubsection{Human-AI Loop in SituFont}
When using SituFont, users can perceive the system's current font adjustments. If the font parameters are not optimal, they can manually adjust the font to better suit the environment. These new adjustments, along with the environmental data, align more closely with the user's personalized reading needs. The Human-AI Loop accelerates data accumulation and updates the model based on user feedback, enhancing the system's adaptability.

\subsection{SituFont System Implementation Overview} \label{subsec:SituFont_implementation}
\begin{figure*}[t]
  \centering
  \includegraphics[width=\textwidth]{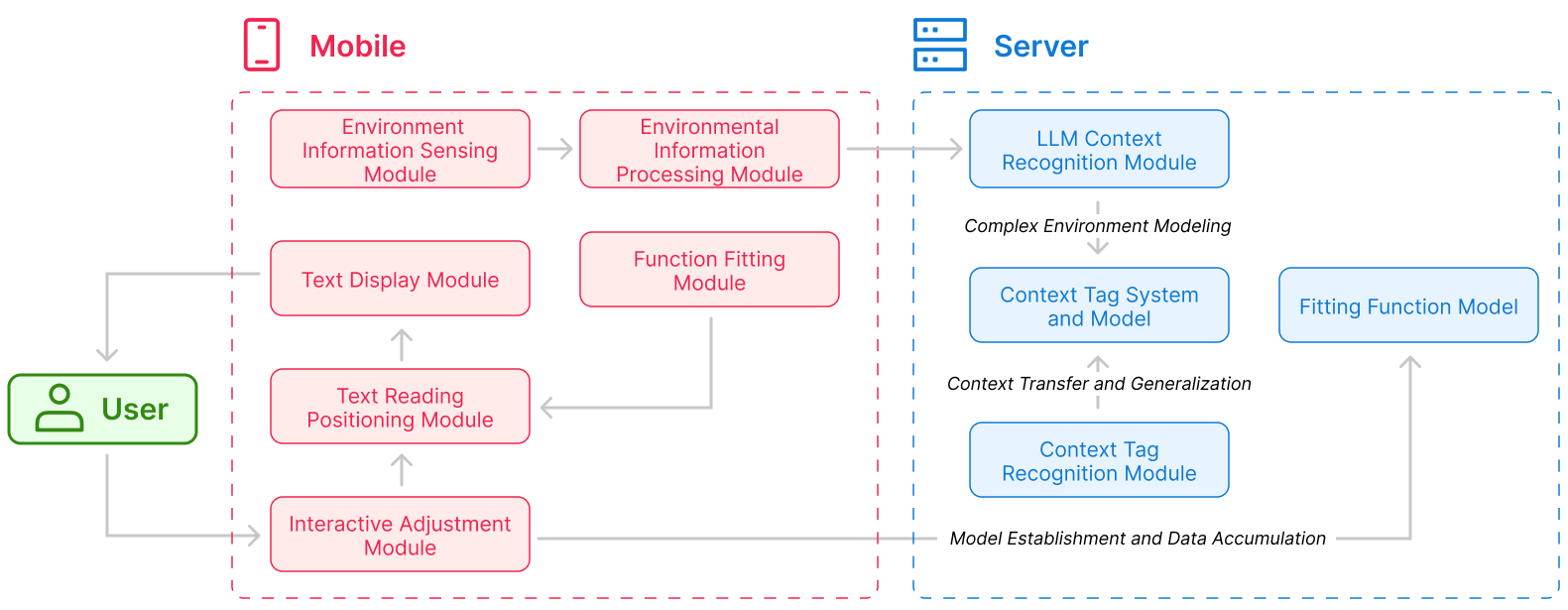}
    \caption{System architecture of SituFont, showing the flow of information from sensors to the adaptation module.}
    \Description{System architecture of SituFont, showing the flow of information from sensors to the adaptation module.}
    \label{fig:system architecture}
\end{figure*}

Based on the system design in Section \ref{subsec:SituFont_design}, we then introduce the implementation details of SituFont. We instantiated SituFont on Android OS (end-user side) and a server (cloud side), as shown in Figure \ref{fig:system architecture}. The SituFont system includes the context sensing module (Section \ref{subsubsec:Context_Sensing}), font adaptation user interface (Section \ref{subsubsec:Adaptive_UI}), the ML pipeline (Section \ref{subsubsec:ML_Pipeline}), and label tree module (Section \ref{subsubsec:Label_Tree}).

\subsubsection{Context Sensing}
\label{subsubsec:Context_Sensing}
The contextual detection system has two key components: context recognition for confirming labels and input for the font adjustment model.

For context recognition and label confirmation, the system utilizes several data sources. It determines location based on GPS POI data and visual input from the rear camera using Vision-Language Models (GPT-4o\footnote{\url{https://openai.com/index/hello-gpt-4o/}}). Additionally, it assesses the user's movement state by analyzing 3-axis vibration data through a pre-trained machine-learning model designed to recognize specific movements. Furthermore, by leveraging Large Language Models  (GPT-3.5Turbo\footnote{\url{https://openai.com/index/gpt-3-5-turbo-fine-tuning-and-api-updates/}}), users can actively describe or modify the current recognized context by typing or using voice input, allowing for more accurate or personalized adjustments to the detected environment.

The font adjustment model relies on various sensor inputs to optimize the reading experience. It monitors ambient light intensity using the mobile phone’s light sensor and takes into account 3-axis vibration data collected from the mobile phone’s accelerometer sensor. The user’s reading distance refers to the distance between their eyes and the screen. To calculate this, MediaPipe’s face recognition functionality is utilized \cite{lugaresi2019mediapipeframeworkbuildingperception}, specifically leveraging the face landmarks to determine the proportion of the eyes in the image, which is then converted into the actual reading distance by factoring in the user's real interpupillary distance (IPD) before detection. A similar method is used in AngleSizer to detect the distance between two hands \cite{Xiaoqing2024}. 

\subsubsection{Adaptive User Interface}
\label{subsubsec:Adaptive_UI}

As described in Section \ref{subsubsec:Design_ML} and illustrated in Figure \ref{fig:User interface}, the controls to adjust the text parameters are developed in native Android, allowing dynamic adjustment of the font parameters based on environmental context.

SituFont’s adaptive user interface is designed to balance automation with user control. The system begins with a stable and familiar typographic presentation~\cite{wang2024adaptive}, which serves as an initial reference point rather than an assumed optimal configuration. By default, the interface displays black text on a white background, with a font size of \textbf{20sp}, a font weight of \textbf{400} (regular), a line spacing of \textbf{1.1 em}, and a letter spacing of \textbf{0.03 em}. These values were selected based on prior work on mobile legibility and established guidelines for Chinese text layout~\cite{huang2019effects, huang2009effects, wang2008chinese}.
These parameters do not change automatically. Instead, typographic adaptation is initiated by the user through a long-press gesture, which triggers the Font Adjustment Model to generate context-sensitive recommendations. A brief vibration confirms the activation, after which the suggested font parameters are applied. This interaction design ensures that adaptation occurs at moments chosen by the user, reducing unintended disruptions while allowing the system to respond effectively to changing reading conditions.

During operation, environmental parameters such as ambient light, reading distance, and device acceleration are continuously sampled at a rate of 5Hz (every 200 milliseconds), but the font remains unchanged unless a long-press gesture is detected. Once triggered, the model infers a new set of font parameters, including size, weight, line spacing, and letter spacing, that are then applied to the \texttt{TextView} in real time using Android’s native UI methods. These changes are applied instantly, ensuring a smooth and responsive user experience.

This user-initiated, context-aware adjustment strategy offers a balance between automation and user control. It prevents intrusive or unnecessary UI changes while allowing the system to optimize readability when needed. Such a design also enhances reproducibility: the adaptation logic is simple to implement with access to environmental sensors and can be replicated using either rule-based logic or ML-based inference, as detailed in Sections~\ref{subsubsec:ML_Pipeline} and~\ref{subsubsec:Label_Tree}.

\subsubsection{ML Pipeline}
\label{subsubsec:ML_Pipeline}
Since SituFont utilizes a simple Regression Machine Learning model with a small amount of fit data, the delay for both training and inference is negligible. The data for training and inference is collected directly on the user's Android mobile phone. Once transmitted to the cloud backend, the model is typically updated or the inference results are returned to the front end within 2 seconds, allowing for prompt font parameter adjustments.

\subsubsection{Label Tree Implementation}
\label{subsubsec:Label_Tree}
The implementation of the Label Tree consists of two main components: Label Generation and Label Selection. After gathering the context information outlined in Section \ref{subsubsec:Context_Sensing}, the system first uses an LLM to select the most appropriate label from the existing labels stored in the cloud database. If no suitable label is found, or the selected label does not match the user’s current context, the system generates a new label based on the Label Tree Structure described in Section \ref{subsubsec:Label_Tree_Design} and the newly provided context information. Once the user confirms the newly generated label, it is stored in the user’s cloud database for future use. The prompts used in Label Generation and Label Selection are listed in Appendix \ref{appendix:Label Tree Prompt}.

\section{User study}
To evaluate SituFont's impact on reading performance and user experience compared to a traditional \revision{reading interface} that requires manual font adjustments, we conducted a within-subject study with 12 participants under eight simulated SVI scenarios. The study consisted of three phases: (1) a pre-test establishing baseline reading performance under eight SVI conditions, (2) a four-day adaptation period using SituFont, and (3) experimental reading tasks comparing SituFont and a traditional \revision{reading interface} under identical conditions (Figure \ref{fig:userstudytimeline}). 

The user study aimed to address two research questions: 

\textbf{RQ1:} Does SituFont improve reading performance compared to traditional displays under varying SVI conditions? 

\textbf{RQ2:} How do users perceive SituFont’s workload and overall experience compared to traditional displays?

\subsection{Participants and Apparatus}
Participants were recruited via public social media posts. 12 participants took part in the study (5 male, 6 female, 1 non-binary, $M = 22.3$, $SD = 4.1$, age range = 18 to 34). All participants were native Mandarin Chinese speakers. Participants reported using a variety of devices for reading, including smartphones ($n = 12$), tablets ($n = 6$), and laptops ($n = 7$). \revision{Daily sustained reading time on smartphones, meaning the reading of longer and more cognitively demanding passages that require comprehension and focused engagement, ranged from 10 minutes to over 2 hours.} Ten participants reported wearing corrective lenses. One participant reported difficulty reading standard-sized text on mobile screens due to eye strain, while another reported general eye fatigue. All participants provided written informed consent.

\revision{All pre-test and experimental sessions used a \revision{HUAWEI P40 (6.1-inch screen, 2340×1080 resolution)} standard Android phone to ensure consistency. During the adaptation period, participants used either their own Android devices or the provided Android phone. }


\subsection{Design}
\begin{figure*}[t]
  \centering
  \includegraphics[width=\textwidth]{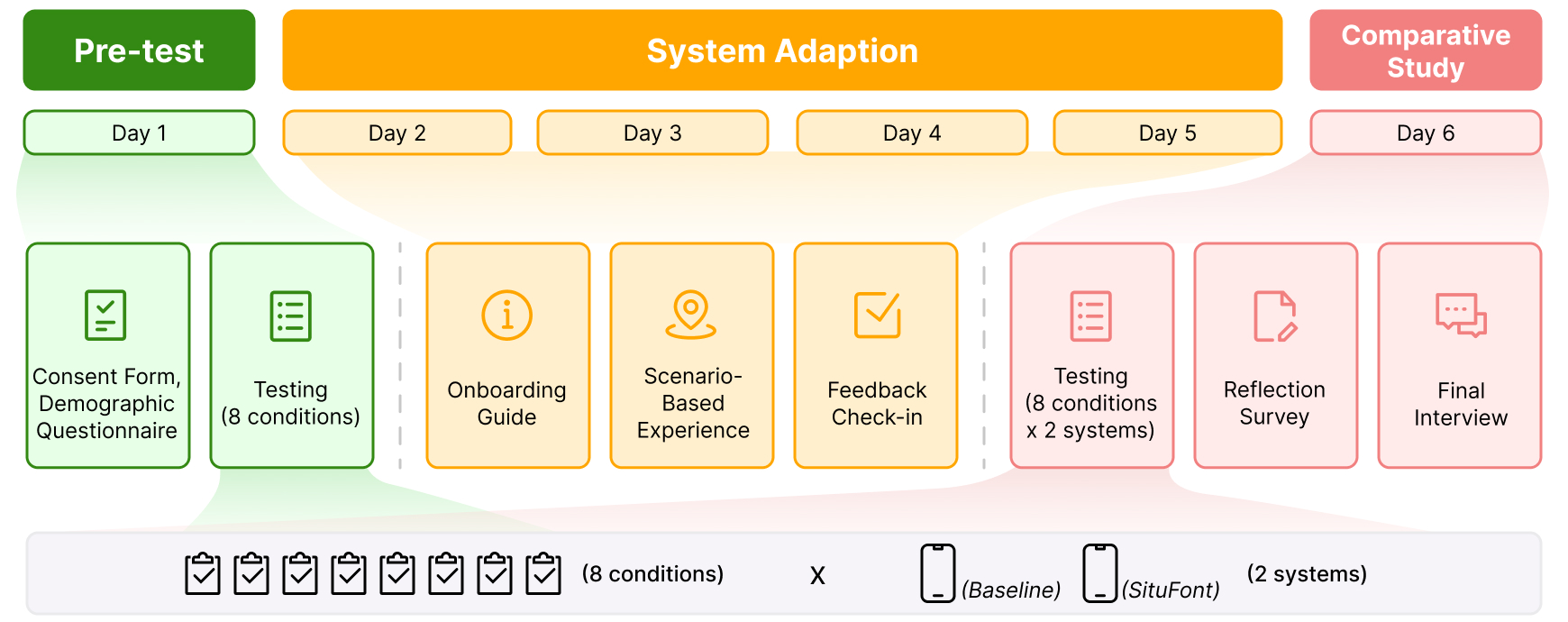}
    \caption{User study timeline.}
    \Description{User study timeline.}
    \label{fig:userstudytimeline}
\end{figure*}

\revision{The overall study design is shown in Fig.~\ref{fig:userstudytimeline} and consists of three phases: a pre-test, a four-day adaptation period, and a comparative experiment, allowing us to examine SituFont's advantages over a traditional reading interface after substantial adaptation. The pre-test serves as a baseline for task completion under the traditional interface, enabling us to assess whether participants exhibit learning effects on the reading tasks themselves by comparing traditional-interface performance between the pre-test and post-test. The primary comparison of interest, however, is between SituFont and the traditional interface in the final phase. All controlled experiments, including the pre-test and both settings of the comparative experiment, were conducted under the eight SVI conditions. In each condition, participants completed two tasks: a read-aloud task and a reading comprehension task. Below, we describe the experimental conditions, corpus, tasks, and baseline configuration.}

\subsubsection{Conditions}
\begin{figure}
    \centering
    \includegraphics[width=1\linewidth]{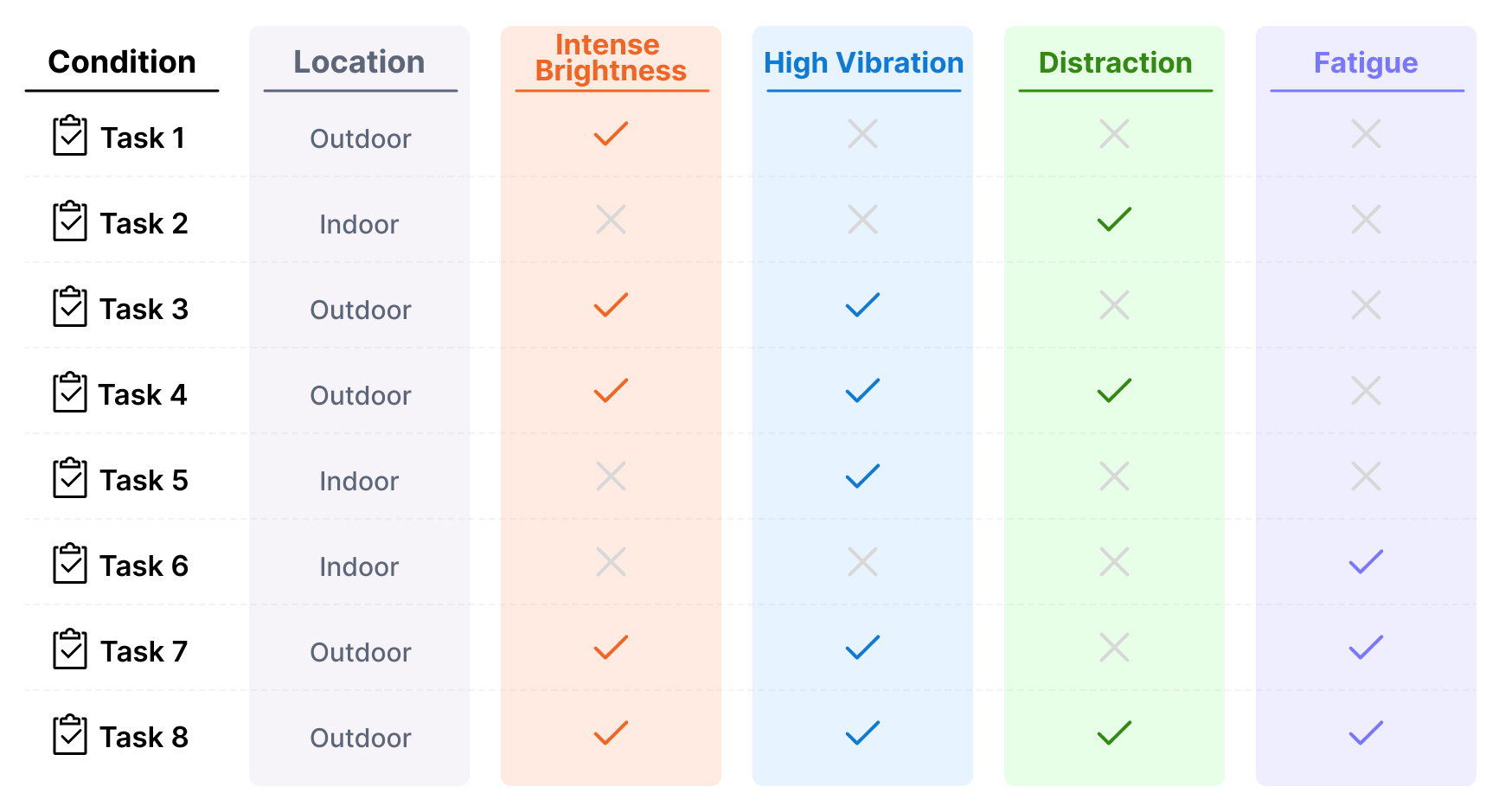}
    \caption{An example order of the eight experiment conditions for comparative study.}
    \Description{An example order of the eight experiment conditions for comparative study.}
    \label{fig:experimentconditions}
\end{figure}

Guided by prior literature on SVIs \cite{tigwell2019addressing, mustonen2004examining, tigwell2018designing} and insights from our formative study, we selected eight experimental conditions (Table \ref{tab:experimental_conditions}, Figure \ref{fig:experimentconditions}) to balance ecological validity with feasibility. We prioritized four primary SVI factors: intense brightness, high vibration, distraction, and fatigue, all frequently cited as impactful in mobile reading contexts. Intense brightness and high vibration were especially disruptive, often occurring outdoors (e.g., reading in sunlight or while moving), while distraction and fatigue were more relevant to indoor settings or prolonged use. Testing all 16 possible combinations was impractical due to participant burden, so we selected a representative subset that reflected the most common and realistic scenarios. For example, \textit{Intense Brightness + High Vibration + Distraction} simulates running in crowded outdoor spaces, and \textit{Intense Brightness + High Vibration + Fatigue} mirrors the compounded strain of outdoor exercise. Scenarios lacking brightness or vibration were excluded, as the formative study indicated they were less frequent and less disruptive in real-world use.

\subsubsection{Tasks and Corpus}

\revision{All experimental sessions involved two tasks: a read-aloud task and a reading comprehension task. The read-aloud task required participants to read the entire passage aloud at a natural and comfortable pace. The reading comprehension task required participants to complete five multiple-choice questions designed to assess their understanding of the passage content. Participants completed these comprehension questions at a natural, self-paced rhythm without time constraints.}

Each condition used two 50-character passages sourced from the HSK Level 5 examination. \revision{Although originally designed for advanced non-native learners, HSK Level 5 provides standardized Chinese passages of medium to high difficulty that remain appropriate for native speakers \cite{yang2025navigating, peng2021hanyu}.} The HSK corpus also includes predefined multiple-choice comprehension questions, ensuring consistent difficulty and minimizing memorization bias \cite{LiYongquan2022}.

\subsubsection{Baseline}

\revision{To evaluate the effectiveness of SituFont, we selected a competitive and realistic baseline: a manually adjustable reading interface that matches SituFont in both interface elements and parameter degrees of freedom. This baseline operates as follows. Before each reading task, under a specific SVI condition, participants were allowed unlimited time to freely adjust all font parameters under the template corpus until they reached what they perceived as the optimal static configuration. \textbf{Adjustment time was not counted as part of the task, and no further adjustments were allowed once reading began.}}

\revision{We argue that such a baseline setting is fair and non-trivial. First, using a fixed-layout baseline would be too weak, as it would substantially underestimate the upper bound of manual optimization and may introduce avoidable readability issues and unfairness for different SVI scenarios, thereby exaggerating the advantages of SituFont. Second, allowing participants to tune the interface before each task yields a near-optimal static layout for the given scenario. This enables two meaningful comparisons: whether SituFont’s predicted parameters approach this near-optimal configuration, and whether SituFont’s ability to adapt dynamically to changing conditions provides additional benefits beyond what static optimization can achieve. Third, this baseline reflects a fair and realistic usage pattern. In everyday reading, users typically adjust settings before reading rather than during the act of reading itself. By excluding adjustment time from task completion and keeping the reading process identical across conditions, we ensure that SituFont and the baseline differ only in their adaptation mechanisms rather than other operational interferences.}

\subsection{Procedures}

\subsubsection{Pre-test}
Before the adaptation period, participants completed a baseline reading pre-test using a traditional mobile reading interface with manual font control, across eight SVI conditions (Table \ref{tab:experimental_conditions}). Comprehension was assessed using five HSK-defined multiple-choice questions per passage, with accuracy calculated as the percentage of correct responses. Reading performance was measured via goodput (characters per minute, CPM), defined as the number of correctly read characters divided by total reading time \cite{Ku2019Peri}. Participants read aloud at their normal pace, with audio recorded for later analysis. A pre-test questionnaire also gathered demographic information, vision history, and reading habits (Appendix \ref{appendix: Pre-test Survey}).

\subsubsection{Adaptation Period}
Over four days, participants used SituFont in three to four different reading scenarios daily. A daily survey captured reading contexts, perceived fatigue, distraction levels, and usability issues (Appendix \ref{appendix: Daily Survey}). This phase allowed participants to \revision{label their personalized preferences, where the prediction model learns personalization adaptation strategies,} and become familiar with the system before the main evaluation.

\subsubsection{Post-Adaptation Reading Evaluation}
Following the adaptation period, participants completed the same reading tasks under the eight SVI conditions, this time using both SituFont and a traditional display. Each condition included two new HSK Level 5 passages. Interface order was counterbalanced across participants and conditions. After each passage, participants answered five standardized comprehension questions and their reading goodput was calculated. They then completed the NASA-TLX questionnaire to assess perceived workload. After all conditions, participants completed UEQ-S and SUS questionnaires and participated in a semi-structured interview (Appendix \ref{appendix: Post-test interview}).

\begin{table*}[h]
\centering
\small
\renewcommand{\arraystretch}{1.2} 
\begin{tabularx}{\textwidth}{lccc}
\toprule
\textbf{Condition} & \textbf{Lighting} & \textbf{Motion} & \textbf{Task Load} \\
\midrule
Intense Brightness & 50,000+ lux\textsuperscript{(a)} & Static & None \\
Distraction & Normal (Indoor) & Walking \textsuperscript{(b)} & Navigation Task\textsuperscript{(c)} \\
High Vibration & Normal (Indoor) & Running\textsuperscript{(d)} & None \\
Fatigue & Normal (Indoor) & Walking\textsuperscript{(b)} & Weighted Load\textsuperscript{(e)} \\
Intense Brightness + High Vibration & 50,000+ lux\textsuperscript{(a)} & Running\textsuperscript{(b)} & None \\
Intense Brightness + High Vibration + Distraction & 50,000+ lux\textsuperscript{(a)} & Running\textsuperscript{(b)} & Navigation Task\textsuperscript{(c)} \\
Intense Brightness + High Vibration + Fatigue & 50,000+ lux\textsuperscript{(a)} & Running\textsuperscript{(b)} & Weighted Load\textsuperscript{(e)} \\
Intense Brightness + High Vibration + Fatigue + Distraction & 50,000+ lux\textsuperscript{(a)} & Running\textsuperscript{(b)} & Navigation Task\textsuperscript{(c)} + Weighted Load\textsuperscript{(e)} \\
\bottomrule
\end{tabularx}
\caption{Experimental Conditions in User Study}
\label{tab:experimental_conditions}
\begin{flushleft}
\textsuperscript{(a)} Outdoor conditions occur under strong midday natural lighting (50,000+ lux, clear sky). \\
\textsuperscript{(b)} Motion occurs along a straight 50m path, either indoors (office hallway, no obstacles) or outdoors (running track, no pedestrians). \\
\textsuperscript{(c)} Participants navigated through soccer training cones (30 cm tall) placed 5 meters apart with no external distractions, passing yellow cones on the left and orange cones on the right while reading.\\
\textsuperscript{(d)} High-vibration treadmill condition involves running at 6 km/h with no incline or handrails. \\
\textsuperscript{(e)} Weighted load consists of a 3 kg backpack (containing a laptop and two books), worn with straps adjusted snugly for even weight distribution. \\
\end{flushleft}
\end{table*}


\subsection{Measures} 

\subsubsection{Reading Performance}  
Reading performance was assessed through goodput (characters per minute, CPM) and reading comprehension accuracy.

Goodput was defined as the number of characters correctly read aloud by the participant, divided by the total reading time, following and adapting the approach from Ku et al.~\cite{Ku2019Peri}. While Ku et al. measured words correctly perceived in a rapid serial visual presentation setting using AR glasses, we extended this to naturalistic read-aloud tasks on mobile devices under SVI conditions. Participants were instructed to read each passage aloud at a comfortable pace during each SVI condition. Reading time was measured from the moment the participant began reading until they completed the passage. A character was considered “correct” if it matched the corresponding character in the source text. This was done while ignoring case, punctuation, and minor mispronunciation. Goodput was calculated as:
\[
\text{Goodput} = \frac{C}{T}
\]
\ where $C$ is the number of correctly read characters, and $T$ is the total reading time in minutes. This metric intends to capture both speed and accuracy. 

Reading comprehension was measured by the percentage of correct responses to five multiple-choice questions per passage. These questions were pre-defined based on the HSK Level 5 Mandarin proficiency examination \cite{barnard2007capturing, Zhou2023Not}.

\subsubsection{Perceived Workload \& User Experience} Perceived workload was evaluated using NASA-TLX \cite{hart1988development}, measuring mental demand, physical demand, temporal demand, performance, effort, and frustration. Overall usability was assessed through UEQ-S \cite{Schrepp2017} and SUS \cite{BangorKortumMiller2008}, capturing users' subjective impressions of each interface.

\subsubsection{Qualitative Feedback}
Post-study semi-structured interviews were conducted to collect in-depth feedback on participants’ experiences with both SituFont and the traditional display, including difficulties encountered, perceived benefits, and interface preferences. Interview transcripts were thematically analyzed following Braun and Clarke’s guidelines \cite{braun2006using}. Two researchers independently performed open coding on the first six transcripts using MAXQDA \cite{maxqda2022}, after which a refined codebook was developed through discussion. This codebook was then applied to the remaining six transcripts. Final themes were generated by clustering related codes and refining them collaboratively.

\section{Result}
Over the four-day adaptation period, participants generated 490 valid data entries, with an average of 7 recorded reading scenarios per participant. The following sections evaluate SituFont's effectiveness in improving reading performance, reducing workload, and enhancing user experience across the eight SVI conditions. Before conducting parametric tests (e.g., paired t-tests, ANOVA), we assessed data normality for each metric using the Shapiro–Wilk test. All variables used in parametric analyses (e.g., goodput, comprehension accuracy, UEQ-S, and SUS) met the normality assumption ($p > 0.05$). For metrics that violated normality (e.g., NASA-TLX workload subscales), we used the Wilcoxon signed-rank test as a non-parametric alternative.

\begin{table*}[h]
\centering
\begin{threeparttable}
\begin{tabular}{lcccccccc}
\toprule
Scenarios ID \footnotemark[1] & 1 & 2 & 3 & 4 & 5 & 6 & 7 & 8 \\
\midrule
Baseline Pre-test & 254.17 & 252.59 & 208.15 & 206.84 & 272.69 & 289.52 & 277.61 & 226.04 \\
& (40.72) & (42.43) & (44.15) & (40.38) & (40.40) & (54.93) & (39.95) & (45.80) \\
SituFont & 300.64 & 288.25 & 271.51&304.58& 295.78&352.52&288.34&292.83\\
 & (56.80) & (40.72) & (50.05)&(39.36)& (48.91)&(51.06)&(31.97)&(47.65)\\
Baseline Post-test & 285.32 & 241.95 & 256.10 & 249.14 & 273.26 & 273.33 & 256.77 & 246.83 \\
& (39.66) & (49.01) & (70.46) & (52.39) & (50.30) & (51.05) & (50.38) & (53.02) \\

\midrule
Situ v.s. Baseline - \textit{T} Value & -1.961 & -7.741 & -2.166 & -5.634 & -2.742 & -10.122 & -2.245 & -3.256 \\
Situ v.s. Baseline - \textit{p} Value & 0.076 & 0.000*** & 0.052 & 0.000*** & 0.019* & 0.000*** & 0.046* & 0.008** \\
\midrule
Pre v.s. Post - \textit{T} Value & -3.36 & 1.29 & -3.09 & -4.31 & 0.04 & 1.14 & 1.40 & -2.73 \\
Pre v.s. Post - \textit{p} Value & 0.006** & 0.22 & 0.01* & 0.001** & 0.97 & 0.28 & 0.19 & 0.02* \\
\bottomrule
\end{tabular}
\begin{tablenotes}[flushleft]
\item[1] For convenience of analysis, scenarios are labeled as:
1 -- Intense Brightness + Normal; 
2 -- Distraction + Normal; 
3 -- Intense Brightness + High Vibration + Normal; 
4 -- Intense Brightness + High Vibration + Distraction; 
5 -- High Vibration + Normal; 
6 -- Fatigue + Normal; 
7 -- Intense Brightness + High Vibration + Fatigue + Normal; 
8 -- Intense Brightness + High Vibration + Distraction + Fatigue.
\item[*] \textit{p} < 0.05, ** \textit{p} < 0.01, *** \textit{p} < 0.001
\end{tablenotes}
\caption{Mean (standard deviation) of reading goodput (CPM) across different SVIs scenarios. The results of paired t-tests demonstrated that SituFont's observed goodput improvement was statistically significant}
\label{tab:goodoutput}
\end{threeparttable}
\end{table*}

\subsection{Reading Goodput}

\revision{Across conditions, SituFont consistently improved users’ reading efficiency, yielding higher goodput scores than the baseline system. Paired sample t-tests confirmed this improvement, showing that, except for the \textit{Intense Brightness} Scenario ($t = -1.961, p = 0.076$) and \textit{High Vibration} Scenario ($t = -2.166, p = 0.052$), SituFont significantly outperformed the Baseline system in all other conditions (Figure \ref{fig:overall_performance}). We interpret these two single-factor results as marginal trends that did not reach the conventional $p < .05$ threshold and are consistent with smaller or context-dependent effects when a single SVI factor is present in isolation. Across scenarios, effect sizes ranged from moderate in the two marginal single-factor cases (\textit{Intense Brightness}: \(d = 0.57\); \textit{High Vibration}: \(d = 0.79\)) to large or very large in multi-factor scenarios (e.g., \textit{Intense Brightness + High Vibration + Distraction}: \(d = 1.63\); \textit{Fatigue}: \(d = 2.92\)).
The largest gains were observed in scenarios involving multiple simultaneous impairments, such as \textit{Intense Brightness + High Vibration + Distraction} ($\Delta=79.2$) and \textit{Intense Brightness + High Vibration + Fatigue + Distraction} ($\Delta=46.0$), suggesting that SituFont’s dynamic adjustments are particularly beneficial when contextual factors interact in complex ways. In contrast, the improvements in the \textit{Intense Brightness} and \textit{High Vibration} single-factor scenarios did not reach statistical significance. This pattern suggests that when the impairment is dominated by a single, strong factor, participants can often approximate a reasonable static configuration during the baseline optimization phase, narrowing the performance gap between static and adaptive adjustments.}

\revision{A comparison between pre-test and post-test performance under the traditional interface further shows that users exhibited some adaptation to the task itself. Goodput increased in several conditions after repeated exposure to similar reading tasks, indicating a modest learning effect. However, this improvement did not eliminate SituFont’s advantage. Even after users had opportunities to learn optimal manual adjustments and become familiar with the reading tasks, SituFont continued to outperform the manually optimized baseline in the majority of SVI conditions. This finding reinforces that SituFont’s benefit stems not only from parameter selection but also from its ability to respond dynamically to fluctuating SVI factors, which a static configuration cannot capture.}

\revision{Together, these results suggest that SituFont provides robust performance advantages across diverse and dynamically changing environments, particularly in scenarios where multiple SVI factors combine to create substantial visual and cognitive challenges.}

\subsection{Reading Comprehension}

\begin{figure*}[t]
  \centering
  \includegraphics[width=\textwidth]{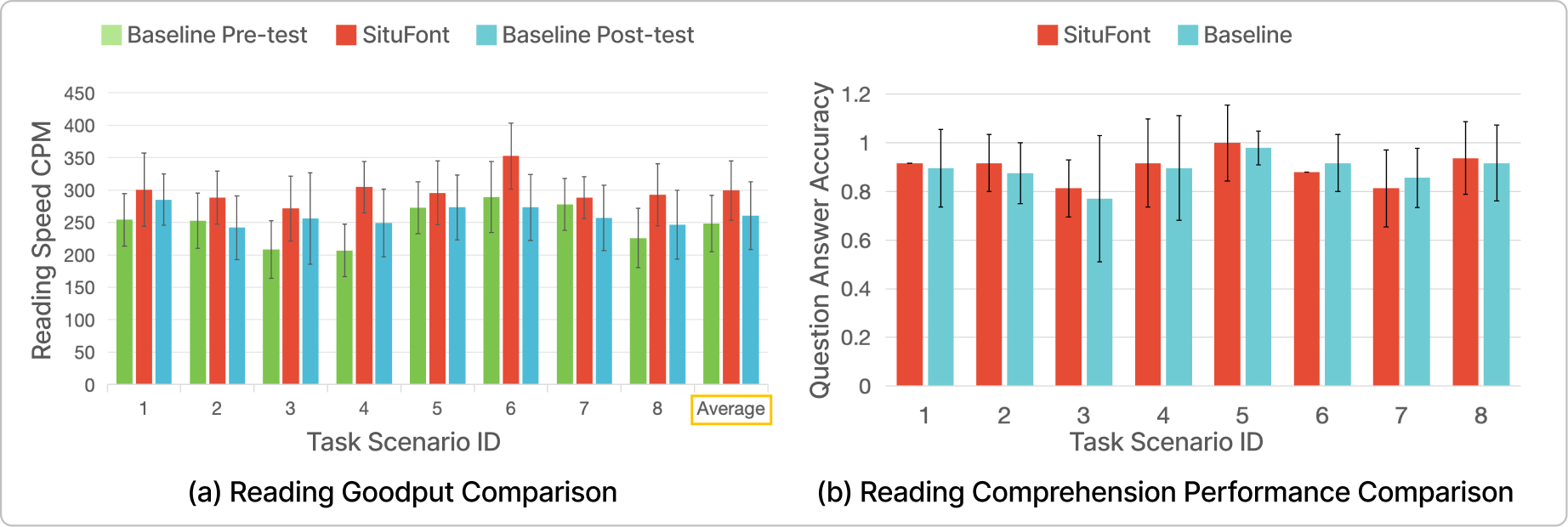}
    \caption{Comparison of reading goodput (a) and comprehension accuracy (b) across different SVI conditions. SituFont consistently demonstrated improved goodput, while comprehension accuracy remained stable across both interfaces.}
    \Description{Comparison of reading goodput (a) and comprehension accuracy (b) across different SVI conditions. SituFont consistently demonstrated improved goodput, while comprehension accuracy remained stable across both interfaces.}
    \label{fig:overall_performance}
    \footnotetext[1]{For the convenience of analysis, scenarios are labeled as: 
1 - Intense Brightness + Normal, 
2 - Distraction + Normal, 
3 - Intense Brightness + High Vib + Normal, 
4 - Intense Brightness + High Vib + Distraction, 
5 - High Vib + Normal, 
6 - Fatigue + Normal, 
7 - Intense Brightness + High Vib + Fatigue + Normal, 
8 - Intense Brightness + High Vib + Distraction + Fatigue.}
\end{figure*}

\revision{Overall, reading comprehension accuracy remained comparable between SituFont and the Baseline system across all task scenarios (Figure \ref{fig:overall_performance}). Although SituFont showed slightly higher mean accuracy than the Baseline system in most conditions (except Scenario 7), none of these differences reached statistical significance. This stability is consistent with our task design, where participants completed comprehension questions in a natural manner, meaning that comprehension performance was driven primarily by text difficulty rather than interface differences. The overlapping error bars further indicate that both interfaces supported a similar level of understanding, suggesting that while SituFont improves reading efficiency, it does not alter users’ ability to comprehend the passage content.}

\subsection{Perceived Workload}

\begin{figure*}[t]
  \centering
  \includegraphics[width=\textwidth]{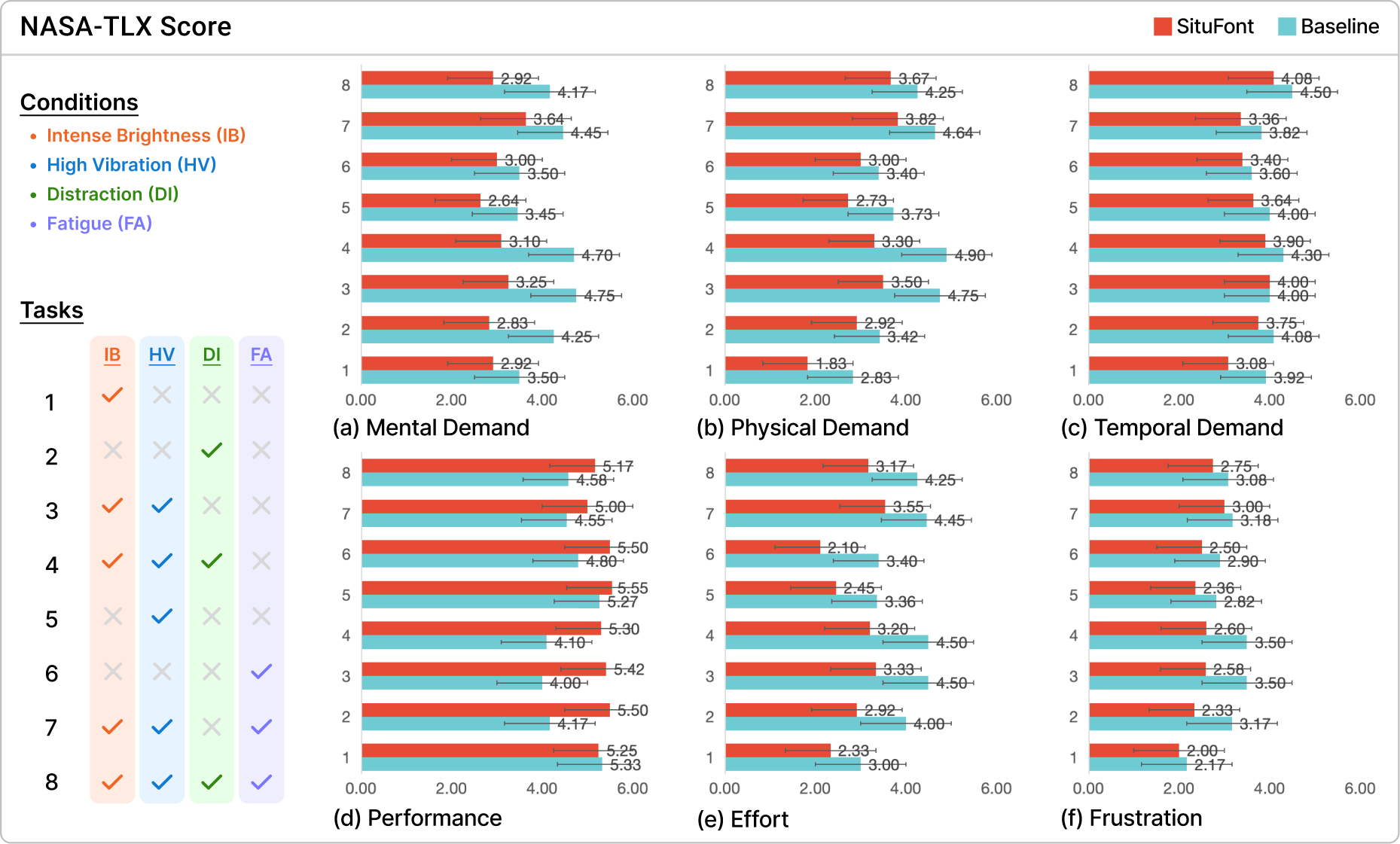}
    \caption{NASA-TLX scores across experimental conditions, illustrating that SituFont consistently reduced mental and physical workload compared to the Baseline system. No significant differences were observed in temporal demand, while frustration levels showed minor variations. For convenience of analysis, we labeled different experimental conditions with numbers, with each number representing specific conditions as shown on the left.}
    \Description{NASA-TLX scores across experimental conditions, illustrating that SituFont consistently reduced mental and physical workload compared to the Baseline system. No significant differences were observed in temporal demand, while frustration levels showed minor variations. For convenience of analysis, we labeled different experimental conditions with numbers, with each number representing specific conditions as shown on the left.}
    \label{fig:NASA-TLX}
\end{figure*}

Participants reported significantly lower mental and physical workload when using SituFont compared to the Baseline system (Figure \ref{fig:NASA-TLX}). Since the Shapiro-Wilk test indicated that workload measures did not follow a normal distribution ($p < 0.05$ for all), Wilcoxon signed-rank tests were used for pairwise comparisons between the two app conditions within each experimental scenario.

SituFont significantly reduced both mental demand and physical demand in multiple conditions. The Wilcoxon signed-rank test showed that SituFont resulted in significantly lower mental demand under the \textit{Intense Brightness} condition ($W = 2.0$, $p = 0.068$), \textit{Distraction} condition ($W = 0.0$, $p = 0.010$), and \textit{Fatigue} condition ($W = 0.0$, $p = 0.011$). Physical demand was significantly lower in \textit{Intense Brightness} condition ($W = 9.0$, $p = 0.055$), \textit{Intense Brightness + High Vibration} condition ($W = 8.5$, $p = 0.050$), and \textit{High Vibration} condition ($W = 0.0$, $p = 0.016$). These results suggest that SituFont effectively reduces cognitive and physical workload under bright lighting, high vibration, and fatigue-related impairments.

Differences in temporal demand were less pronounced. A statistically significant reduction in temporal demand was found only in the \textit{High Vibration} condition ($W = 9.0$, $p = 0.380$), indicating that SituFont marginally improved time efficiency under vibration stress.

Users perceived higher performance with SituFont under \textit{Distraction} condition ($W = 0.0$, $p = 0.004$), \textit{Intense Brightness + High Vibration} condition ($W = 0.0$, $p = 0.004$), and \textit{Intense Brightness + High Vibration + Distraction} condition ($W = 1.5$, $p = 0.058$). Additionally, effort was significantly lower in \textit{Distraction} condition ($W = 5.5$, $p = 0.041$), \textit{Intense Brightness + High Vibration + Distraction} condition ($W = 0.0$, $p = 0.026$), and \textit{Fatigue} condition ($W = 0.0$, $p = 0.011$). These findings indicate that SituFont enhances reading performance under high-intensity brightness and vibration while reducing effort in distracting and fatigue-inducing environments.

SituFont significantly reduced frustration levels in \textit{Distraction} condition ($W = 5.5$, $p = 0.075$) and \textit{Intense Brightness + High Vibration + Distraction + Fatigue} condition ($W = 6.0$, $p = 0.084$). This suggests that the Baseline system may cause greater frustration under distracting and multi-factor impairment conditions.

\subsection{User Experience and Preferences}

\begin{figure*}[t]
  \centering
  \includegraphics[width=\textwidth]{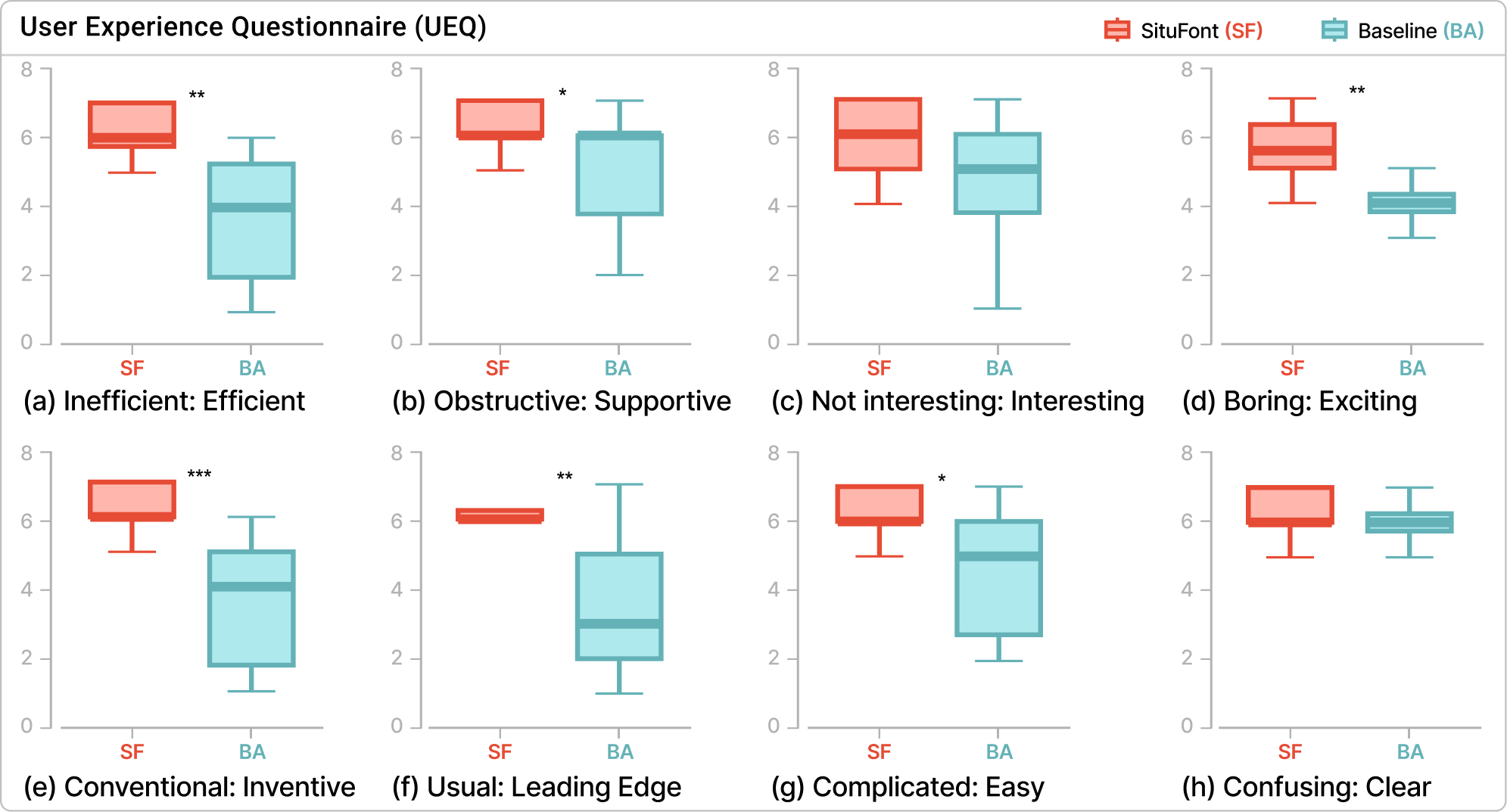}
    \caption{User Experience Questionnaire (UEQ) results, comparing SituFont and the Baseline system across multiple dimensions. SituFont scored significantly higher in efficiency, supportiveness, and novelty, with marginal differences in stimulation and perspicuity. ***: $p$<.001, **: $p$<.01, *: $p$<.05}
    \Description{User Experience Questionnaire (UEQ) results, comparing SituFont and the Baseline system across multiple dimensions. SituFont scored significantly higher in efficiency, supportiveness, and novelty, with marginal differences in stimulation and perspicuity. ***: $p$<.001, **: $p$<.01, *: $p$<.05}
    \label{fig:UEQ}
\end{figure*}

User experience evaluations revealed that SituFont was generally preferred over the Baseline system, with higher ratings in efficiency, ease of understanding, and novelty. While some aspects of stimulation showed marginal differences, participants perceived SituFont as significantly more supportive, efficient, and easy to use (Figure \ref{fig:UEQ}). 

\subsubsection{User Experience Questionnaire (UEQ)}

Attractiveness and Efficiency were key aspects where SituFont demonstrated a clear advantage. In the \textit{Inefficient: Efficient} scale, SituFont scored significantly higher (\textit{M} = 6.08, \textit{SD} = 1.00) compared to the Baseline system (\textit{M} = 3.92, \textit{SD} = 2.27), $p = 0.0005$, suggesting a strong perception of efficiency. Similarly, on the \textit{Obstructive: Supportive} scale, SituFont (\textit{M} = 6.17, \textit{SD} = 0.94) significantly outperformed the Baseline system (\textit{M} = 4.92, \textit{SD} = 1.73), $p = 0.039$, indicating that users found SituFont more supportive in facilitating their reading experience.

The Stimulation scale measured engagement and interest. While SituFont scored higher in \textit{Not interesting: Interesting} (\textit{M} = 5.75, \textit{SD} = 1.36) and \textit{Boring: Exciting} (\textit{M} = 5.50, \textit{SD} = 1.38) compared to the Baseline system (\textit{M} = 4.50, \textit{SD} = 1.83) and (\textit{M} = 4.58, \textit{SD} = 1.88), respectively, the differences did not reach statistical significance ($p = 0.063$ and $p = 0.160$).

Novelty was another category where SituFont was rated higher, indicating a perception of innovation. On the \textit{Conventional: Inventive} scale, SituFont (\textit{M} = 5.58, \textit{SD} = 1.56) significantly outperformed the Baseline system (\textit{M} = 4.08, \textit{SD} = 2.23), $p = 0.024$. Similarly, in the \textit{Usual: Leading edge} dimension, SituFont (\textit{M} = 5.92, \textit{SD} = 1.62) scored significantly higher than the Baseline system (\textit{M} = 3.50, \textit{SD} = 2.11), $p = 0.002$, reinforcing the perception that SituFont introduces an innovative and modern approach.

Perspicuity, which measures ease of understanding, also showed notable differences. SituFont scored a mean of (\textit{M} = 6.25, \textit{SD} = 0.75) for \textit{Confusing: Clear} and (\textit{M} = 6.08, \textit{SD} = 0.90) for \textit{Complicated: Easy}, while the Baseline system scored (\textit{M} = 6.00, \textit{SD} = 0.74) and (\textit{M} = 4.50, \textit{SD} = 1.98), respectively. Although the difference in \textit{Confusing: Clear} was not significant ($p = 0.421$), the difference in \textit{Complicated: Easy} was statistically significant ($p = 0.019$), suggesting that SituFont was perceived as easier to use.

\subsubsection{System Usability Scale (SUS)}

\begin{figure*}[t]
  \centering
  \includegraphics[width=\textwidth]{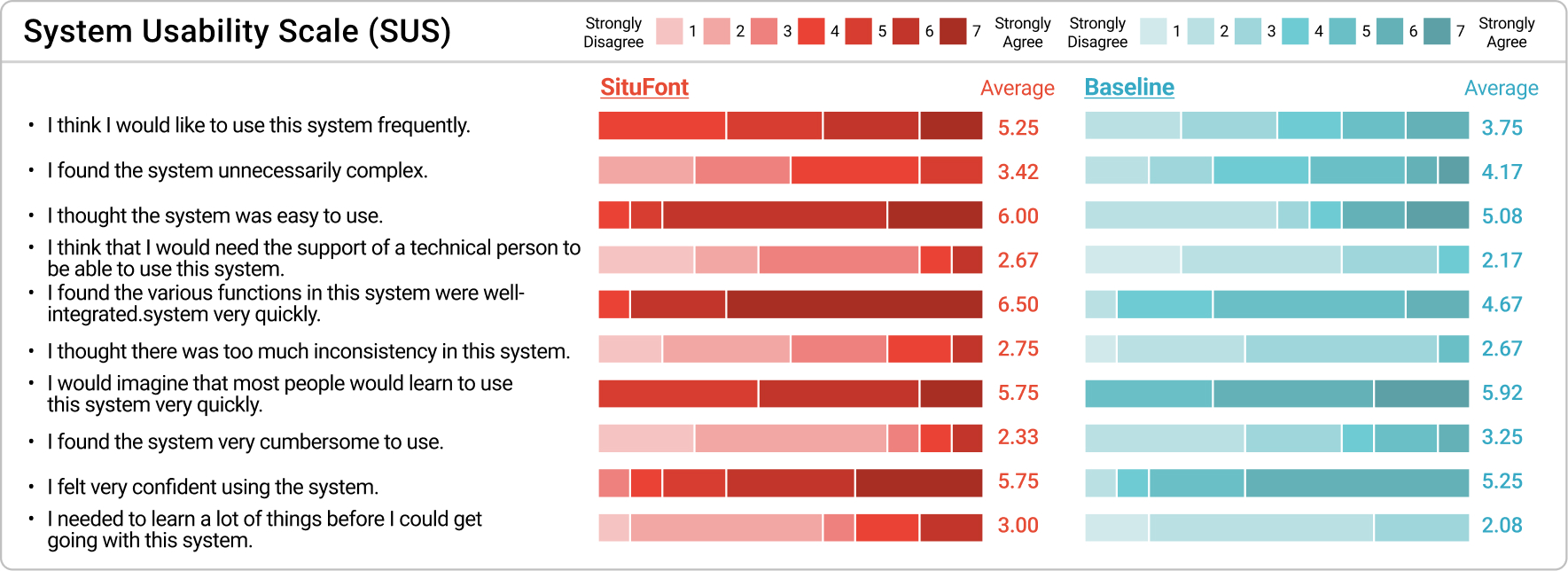}
\caption{System Usability Scale (SUS) results. SituFont was rated significantly lower in complexity and higher in ease of use and consistency, though it required a slightly higher initial learning effort.}
\Description{System Usability Scale (SUS) results. SituFont was rated significantly lower in complexity and higher in ease of use and consistency, though it required a slightly higher initial learning effort.}
\label{fig:SUS}
\end{figure*}

System usability evaluations indicated that SituFont was generally perceived as more user-friendly and consistent, with some minor learning challenges reported (Figure \ref{fig:SUS}).

For the statement \textit{"I found the system unnecessarily complex,"} SituFont (\textit{M} = 2.58, \textit{SD} = 1.62) was rated significantly lower than the Baseline system (\textit{M} = 4.08, \textit{SD} = 2.50), $p = 0.045$, indicating that users perceived SituFont as less complex.

In terms of ease of use, SituFont scored significantly higher on the statement \textit{"I thought the system was easy to use"} (\textit{M} = 6.42, \textit{SD} = 0.90) compared to the Baseline system (\textit{M} = 4.67, \textit{SD} = 2.43), $p = 0.012$, reinforcing that users found SituFont easier to interact with.

For system consistency, SituFont received a mean score of (\textit{M} = 1.08, \textit{SD} = 0.29) for \textit{"I thought there was too much inconsistency in this system"}, significantly lower than the Baseline system (\textit{M} = 1.83, \textit{SD} = 1.03), $p = 0.014$, suggesting that users found SituFont to be more consistent in its functionality.

Confidence in using the system was also measured, with SituFont scoring (\textit{M} = 5.75, \textit{SD} = 1.29) and the Baseline system scoring (\textit{M} = 5.25, \textit{SD} = 1.22), but this difference was not statistically significant ($p = 0.402$).

Lastly, for \textit{"I needed to learn a lot of things before I could use this system,"} SituFont scored higher (\textit{M} = 3.00, \textit{SD} = 1.65) than the Baseline system (\textit{M} = 2.08, \textit{SD} = 0.67), $p = 0.030$, indicating that SituFont required a slightly higher initial learning effort.

\subsubsection{Qualitative Feedback}
Participants valued the convenience of automatic font adjustment, particularly in dynamic contexts such as cycling or running. P2 noted that the system became especially useful once sufficient training data had accumulated, and P5 remarked, ``\textit{When I'm running, long-pressing is much more convenient than manually selecting.}'' The hands-free long-press activation was widely appreciated: P3 described it as effortless, and P4 highlighted its single-handed operation and immersive reading experience. Several participants also reported increased awareness of reading environments; as P9 reflected, ``\textit{It helped me consciously change bad reading scenarios; without this experiment, I wouldn't have realized the impact of reading fonts on the reading experience.}''

Despite these benefits, participants identified several usability issues. Automatic scene detection occasionally misfired due to its reliance on ``\textit{historically recorded descriptions}'' (P8), and long-press activation sometimes lacked precision, with P9 noting that it was ``\textit{sometimes not sensitive enough}'' and other times triggered unintentionally. Manual adjustments via the horizontal slider were challenging in motion: P2 found it difficult to use ``\textit{when the phone is very bumpy},'' and P3 described the slider placement as awkward for one-handed use. Participants suggested improvements such as adding paragraph spacing (P5), reducing disruption from font changes that ``\textit{interrupt the original reading rhythm}'' (P4), and making adaptation more proactive after calibration, particularly for users with larger font needs (P2). Finally, the post-adjustment vibration was perceived as intrusive; P5 commented that it ``\textit{interferes with my reading experience.}''

\section{Discussion and Limitation}

\revision{In this section, we discuss design implications for SVI interventions and JITAI systems with human-in-the-loop feedback, and reflect on ethical and practical considerations. We also discuss limitations and potential future research directions of SituFont.}

\subsection{Design Considerations for SVIs Intervention}

\revision{Our user study compared SituFont with a realistic baseline where participants could freely tune all font parameters before each task until they reached a preferred static configuration, with adjustment time excluded from performance measures. Under these conditions, SituFont still led to higher goodput in most of the eight SVI scenarios, especially when intense brightness and high vibration co-occurred, while comprehension accuracy remained similar. These results suggest that adaptive typography adds value beyond one-off manual tuning by supporting perceptual and cognitive processes that are challenged under SVIs. Prior work on visual span~\cite{luke2024perceptual}, crowding~\cite{schwetlick2025visual, whitney2011visual}, and contrast sensitivity~\cite{maniglia2018effect, vukich2024clinical} suggests that factors such as motion, glare, and vibration can reduce effective perceptual span and increase visual noise, making text harder to process efficiently. SituFont’s context-sensitive adjustments to font size, weight, and spacing help mitigate these perceptual constraints as situational conditions fluctuate, without requiring users to continuously anticipate or correct readability breakdowns. Although participants were able to learn effective manual adjustments over time, doing so required sustained attention and frequent context switching~\cite{lin2024different}, particularly in multi-factor conditions such as walking under intense brightness or vibration. SituFont’s just-in-time adaptations reduced the need for repeated manual intervention, allowing participants to maintain reading flow under changing conditions. }

\revision{The results also show that the success of SVI interventions depends not only on “what” parameters are chosen but also on “how” adaptation is triggered and applied. First, activation must be precise and predictable. From a cognitive perspective, mobile reading under SVIs often involves divided attention~\cite{lin2024different, kong2025supporting, bai2024heads}, as readers must allocate resources between text processing and environmental monitoring~\cite{wang2024safety, yadav2024modeling}. Frequent interruptions or unintended activations can therefore impose additional attentional costs~\cite{chen2024effect, ulku2025eeg}. Participants reported that the current long-press gesture often fired unintentionally during scrolling or text selection, turning adaptation into an interruption. To support SVI scenarios, activation should be “disruption-safe”, input mechanisms that require deliberate intent and include basic safeguards (e.g., suppressing activation when finger velocity is high or when selection handles are active). A vertical edge slider can provide a more ergonomic, deliberate control for one-handed use ~\cite{karlson2006understanding, park2010one, esteves2022one}, and allowing users to customize gestures ~\cite{oh2013challenges} further aligns activation with individual reading habits. Voice commands can offer hands-free control in motion-heavy contexts ~\cite{vu2023voicify, vu2024gptvoicetasker}, but must be balanced against social and privacy concerns.}

\revision{Second, adaptation must preserve layout stability. Sudden jumps in font size or spacing caused participants to lose their place, particularly under high motion. This observation aligns with perceptual accounts emphasizing the role of spatial continuity and stable reference frames in efficient reading~\cite{jensen2025context}. Rather than updating typography immediately with every sensor change, adaptations should align with readers’ micro-rhythms, brief pauses in scrolling or task transitions, and be applied in modest, predictable steps. Simple buffering, such as aggregating small fluctuations into a single adjustment after a short pause, can maintain spatial continuity while still responding to SVIs. In practice, sensing should be treated as a signal that adaptation might be needed; whether and how to adapt should be coordinated with the ongoing reading flow.}

\subsection{Implications on SituFont towards the JITAI Paradigm}

SituFont enables personalized font adaptation by incorporating user feedback, but several aspects require refinement to better capture individual needs. The current binary feedback mechanism (yes/no) is helpful for initial adaptation but lacks the granularity needed for truly personalized adjustments. A more detailed system, such as Likert-scale ratings for distraction, fatigue, and visual strain, or predefined reasons for discomfort, would yield richer data for more tailored interventions. \revision{Participants sometimes responded to a font change even though their discomfort began earlier or was caused by a different environmental factor, making it difficult for the system to interpret which condition the feedback referred to. This temporal mismatch suggests that JITAIs should explicitly capture “when” feedback refers to, such as by storing short interaction histories and allowing users to tag which moment or condition prompted their response. }

Additionally, the cold-start problem limits early system performance due to insufficient user data during initial use, which some participants noted as a period when the system felt less helpful.  This could be mitigated by pre-training on data from Study 2 (\ref{subsection: Study 2}) and fine-tuning with incoming user inputs. Few-shot learning techniques \cite{GongLabel2023,  song2023comprehensive} may further enhance rapid personalization with minimal data. \revision{Beyond model-level strategies, the cold-start period also requires interaction-level scaffolding. For example, early in the deployment, the system could: (a) present users with brief “initial preference snapshots” (e.g., selecting preferred base font size or sensitivity to brightness), (b) temporarily increase explanation frequency (“We enlarged spacing because ambient motion increased”), or (c) provide optional confirmation prompts to reduce miscalibrated early moves. These measures help bootstrap a reliable personalization trajectory while avoiding user frustration during the model’s least accurate phase.}

Long-term usage patterns also merit investigation. Although our study focused on short-term use, participants reported increased awareness of how font settings affect readability ~\cite{wallace2022towards, palmen2023bold}. This learning effect, along with improved reading goodput observed in the Baseline system after exposure to SituFont, suggests that adaptive typography may foster lasting improvements in digital reading strategies. This behavioral shift highlights an important distinction for JITAIs: interventions can serve both an immediate corrective function and a long-term educational function. Systems that recognize this dual role could incorporate mechanisms that gradually decrease intervention frequency as users develop stronger metacognitive control, or allow users to transition between “assistive,” “collaborative,” and “autonomous” modes depending on their evolving confidence and preferences. This pattern also resonates with perspectives on metacognitive awareness and self-regulation in reading, indicating that adaptive typography can function not only as an assistive tool for mitigating SVIs in the moment but also as a scaffold that helps readers learn to manage challenging visual contexts over time. Thus, longitudinal studies should examine whether user behavior continues to evolve and how personalization mechanisms should adapt over extended use.

\subsection{Ethical Considerations}
\revision{SituFont relies on continuous sensing of ambient light and motion and, in some configurations, on external APIs. Several participants questioned permissions and possible “always-on” sensing, echoing broader concerns about privacy and pervasive data collection. Ethical deployment, therefore, requires privacy by design, prioritizing on-device processing and inference, minimizing retention of raw sensor data, and clearly explaining what is collected and how it informs font adjustments. When cloud servers are used for real-time machine learning, techniques such as encryption, federated learning, and differential privacy should be considered to reduce the risk of data breaches ~\cite{hard2018federated, nguyen2016collecting}. Designers must also address potential exclusion, bias, and discriminatory inferences that can arise from cameras, microphones, and large language models~\cite{kumar2025no}, and should recognize that opaque adaptation may undermine trust.}

\revision{Meanwhile, we did not measure battery, latency, or monetary cost in our studies, but continuous sensing and model updates inevitably consume resources. These costs affect both user acceptance and who can benefit from such systems. Providing users with control over sensing frequency and adaptation intensity, as well as simple ways to pause or suspend adaptation, can help negotiate trade-offs between readability, privacy, and energy use. More transparent information about data practices and system behavior, as suggested in work on explainable and trustworthy interfaces~\cite{zerilli2022transparency, wanner2022effect}, may further foster informed adoption. Our findings show that human-in-the-loop adaptive typography can improve reading efficiency and reduce workload under SVIs; future deployments must ensure that these benefits are delivered in a way that remains transparent, controllable, and sustainable for everyday use.}


\subsection{Limitations and Future Work}
Our studies involved young adult native Chinese speakers reading short HSK Level 5 passages on smartphones. As a result, we do not claim generalizability beyond this population and language setting. The HSK materials may also have been insufficiently challenging for some participants, which likely contributed to ceiling effects in comprehension. Future work should recruit older adults and readers with diagnosed visual impairments, and include more demanding and diverse reading tasks (e.g., academic or long-form content) to better probe comprehension and cognitive load under SVIs. 

In addition, we evaluated eight ecologically common SVI conditions rather than all 16 possible combinations and did not include a non-SVI baseline. Scenarios such as \textit{Distraction + Fatigue} without brightness or motion may still meaningfully affect reading. A more complete factorial design with a neutral baseline would help quantify interaction effects and express improvements relative to comfortable reading conditions. We reported only basic accuracy measures for context recognition and did not systematically compare the Label Tree hierarchy to simpler rule-based or flat-feature baselines. Nor did we evaluate feature importance or users’ perceived appropriateness of model outputs. A more complete ML evaluation, including ablation studies and user-facing assessments of perceived fit, would strengthen the justification for our architectural choices and clarify when additional model complexity is warranted.

The current personalization mechanism also has limitations, such as binary comfort feedback being coarse, sometimes temporally ambiguous, and contributing to a cold-start period in the first days of use. Future versions should explore richer yet lightweight feedback structures (e.g., quick tags such as “too small” or “too dense”) and temporally anchored annotations that tie corrections to recent context windows. We also did not evaluate the optional voice-based context channel included in the design, leaving its usability and value an open question. 

Finally, we did not directly measure privacy, energy, or monetary costs. SituFont relies on continuous sensing of ambient light and motion and, in some configurations, external APIs. These channels are important for adaptation but raise privacy concerns and consume battery and data. Future work should investigate privacy-preserving and resource-aware deployments, such as on-device processing, adjustable sensing frequency, user controls over adaptation intensity, and clear communication of data practices. Longer-term field studies will also be needed to examine how personalization strategies evolve over time, whether users continue to rely on adaptation once they internalize effective font settings, and how adaptive typography can be extended to other scripts and accessibility features such as contrast adjustments and screen reader support.

\section{Conclusion}

\revision{Rather than introducing a fundamentally new adaptive framework, this paper builds on existing accessibility and adaptive typography research to show how just-in-time, human-in-the-loop adaptation can support mobile reading under dynamic situational visual impairments through SituFont, and to articulate design considerations grounded in empirical evidence.} Findings from our formative (N=15) and exploratory (N=18) studies identified key SVI factors affecting readability, informing the integration of contextual cues such as motion, lighting, and user preferences into SituFont’s design. Using smartphone sensors and a human-in-the-loop approach, the system dynamically personalizes text presentation to adapt to changing reading conditions. A comparative evaluation (N=12) demonstrated SituFont’s effectiveness in improving readability and reducing task workload across simulated SVI scenarios. \revision{We believe that these insights can inform a broader class of adaptive systems that respond intelligently to environmental and user-centered variability.}

\begin{acks}
This work is supported by the Natural Science Foundation of China under Grant No. 62502411. 
This work is also supported by the Natural Science Foundation of China under Grant No. 62132010, Key Research and Development Program of Ningbo City under Grant No. 2023Z062, Institute for Artificial Intelligence, Tsinghua University (THUAI).
\end{acks}

\bibliographystyle{ACM-Reference-Format}  
\bibliography{situfont}

\appendix
\section{Study 1 Interview Participant Demographics and Interview Protocol}
This appendix provides the participants' demographic and semi-structured interview guide used in Formative Study 1.

\label{appendix: Study 1}
\begin{table*}[t]
\centering
\begin{threeparttable}
\small 
\begin{tabular}{lllll}
\toprule
ID & Gender & Age & Occupation & Vision Status \\ 
\midrule
P1  & Male   & 52  & Public Servant    & Myopia 300, Presbyopia 50, wears corrective glasses \\
P2  & Female & 50  & University Teacher & Presbyopia 100, does not wear glasses daily, wears reading glasses for screens \\
P3  & Female & 46  & Bank Employee     & Normal vision \\
P4  & Male   & 39  & Engineer          & Myopia 300, wears corrective glasses \\
P5  & Male   & 30  & Programmer        & Myopia 600, wears corrective glasses \\
P6  & Female & 28  & University Teacher & Myopia 200, wears corrective glasses \\
P7  & Female & 24  & Bank Employee     & Myopia 200, Astigmatism 150, wears corrective glasses \\
P8  & Male   & 24  & Grad Student      & Myopia 500, wears corrective glasses \\
P9  & Male   & 23  & Grad Student  & Normal vision \\
P10 & Female & 22  & Grad Student  & Myopia 450, wears corrective glasses \\
P11 & Female & 21  & Undergraduate     & Myopia 300, wears corrective glasses \\
P12 & Female & 21  & Grad Student  & Myopia 375, Astigmatism 200, wears corrective glasses \\
P13 & Male   & 20  & Undergraduate     & Normal vision \\
P14 & Male   & 19  & Undergraduate     & Myopia 600, wears corrective glasses \\
P15 & Female & 19  & Undergraduate     & Normal vision \\
\bottomrule
\end{tabular}
\caption{Demographic Overview of Interview Participants}
\label{tab:participants}
\end{threeparttable}
\end{table*}

\textbf{Part 1: Exploring Typical Reading Impairment Scenarios (15–20 minutes)}\\
1. Can you describe a situation where you found it difficult to read on your phone?\\
2. What were you doing at that time? (e.g., walking, riding a bus, lying down)\\
3. What type of content were you trying to read? (e.g., articles, messages, navigation)\\
4. What environmental factors made reading difficult? (e.g., lighting, movement, distractions)\\
5. Did the content’s presentation (e.g., layout, font, content density) contribute to the difficulty?\\

\textbf{Part 2: Current Coping Strategies and Motivations (15 minutes)}\\
6. What do you find most frustrating about reading under difficult conditions?\\
7. How do you typically deal with reading difficulties on your phone?\\
8. Have you tried changing settings like font size or brightness? Why or why not?\\
9. How effective do you find these coping strategies?\\
10. What keeps you motivated to continue reading even when it is challenging?\\

\textbf{Part 3: Concepts for Comfortable Reading Methods (15–20 minutes)}\\
11. In an ideal situation, how could your phone help you read more comfortably?\\
12. What features or changes would make reading easier for you?\\
13. Would you prefer the system to adjust settings automatically, or would you like manual control?\\
14. How would you like the system to interact with you? (e.g., quietly in the background, notify you, ask for input)\\
15. What would make you feel comfortable using such a system?\\

\section{Study 2 Experimental System and Equipment}
\label{appendix:Experiment}
We developed an application with six preloaded reading materials in Chinese, ensuring isomorphic language, structure, and difficulty (high school level). Each 550-character passage was presented in plain text with no additional visual elements. Passages were preloaded, each appearing once during the experiment. The application allowed real-time text adjustments via a one-handed interface: double-tapping opened a menu for font size, weight, line spacing, and letter spacing. Smartphone sensors recorded light intensity, three-axis acceleration, and reading distance. After participants completed their adjustments, they could upload the text parameter data and environmental data by tapping any other area on the screen. To avoid the impact of screen size and resolution differences, we used two HUAWEI P40 smartphones as the experimental devices, each with a 6.1-inch screen and a resolution of 2340×1080 pixels.

\textit{Experimental Procedure.}
Before the experiment, the lead researcher assigned participant IDs, explained procedures, and demonstrated a trial. Participants read passages in indoor and outdoor settings while standing, walking, and running, with 3-minute reading durations per passage. During reading, participants encountered SVIs caused by environmental factors such as strong light, vibration, or reading distance, and they were instructed to adjust text parameters to mitigate these issues. Devices were held one-handed, with adjustments permitted anytime.

\textit{Data Collection}
Smartphone sensors recorded environmental data throughout the experiment. The reading distance was measured using the front camera, the intensity of the ambient light was captured by the light sensor, and the vibration displacement of the three axes was tracked by the accelerometer. The adjustments of the text parameters were recorded using the built-in functions of Android, with line spacing and letter spacing expressed in \textit{em} units. For example, a line spacing of 1\textit{em} equaled the text height, while 0.05\textit{em} represented 5\% of the font size.

\section{Label Tree Prompt}
\label{appendix:Label Tree Prompt}
\subsection{Context Label Generation \& Selection}
\textbf{system\_role}: \\
(You need to determine my current status based on the environmental information and photos I provide, including movement, environment, and personalized description. \\
\begin{itemize}
  \item \textit{Movement} refers to my current physical activity, 
  \item \textit{Environment} refers to the setting I am currently in, and 
  \item \textit{Personalized Description} refers to any specific details related to my status or environment. 
\end{itemize}
In order to obtain labels that describe my status, you should follow the steps below based on the environmental information and two photos I provide:)

\textbf{step1}: \\
(\textit{Step 1: Based on the location information, movement status, and photos I provide, choose my 'movement status' from the following three options: 1. Still 2. Walking 3. Running})

\textbf{step2}: \\
(\textit{Step 2: Combine the provided location information and images to determine my current 'environment name.'})

\textbf{step3}: \\
(\textit{Step 3: Based on the judgments from the first two steps and the photos, select the most appropriate status label from the provided label options:})

\textbf{example}: \\
(\textit{For example: If my location and photos show that I am in an office, and my movement status is still, the status label would be "Still - Office";})

\textbf{attention}: \\
(\textit{Note: Ensure the result format is 'Movement-Environment-Personalized Description' or 'Movement-Environment.' \\
!!! Please strictly return results in the following format:} \\
\textbf{Movement-Environment-Personalized Description} or \textbf{Movement-Environment})

\subsection{Prompt for Editing Label}
\textbf{user\_description\_prompt}: \\
(\textit{This is the current label used to describe my Movement-Environment-Personalized Description:})

\textbf{user\_step1}: \\
(\textit{I first want to modify some parts of the movement, environment, or personalized description. \\
For example, if the user says, "I am in an office," the user wants to change the environment. \\
If the user says, "I am running," then the user wants to modify the movement. \\
If the user says, "I am wearing a hat," the personalized description should add "wearing a hat." \\
If the user says, "I am not wearing glasses," the personalized description should emphasize "no glasses."})

\textbf{user\_step2}: \\
(\textit{Based on the following requirements, update the current Movement-Environment-Personalized Description to reflect the new status.})

\textbf{format\_attention}: \\
(\textit{!!! Remember: only output the modified label, nothing else.})

\textbf{select\_label\_role}: \\
(\textit{You need to select the closest matching label from an existing label list based on the environmental information, photos, and current status label I provide.})

\section{User Study Questionnaires and Interview Protocol}
\label{appendix:UserStudy}

This appendix provides the full set of questionnaires and the interview protocol used in the user study. Standardized scales (NASA-TLX, UEQ-S, and SUS) were used without modifications and are not included here. Participants completed these surveys to provide demographic information, context data during the adaptation period, and qualitative feedback in a post-study interview.

\subsection{Pre-test Demographic and Vision Questionnaire}
\label{appendix: Pre-test Survey}
1. Age: a. 18–24 b. 25–34 c. 35–44 d. 45–54 e. 55–64 f. 65+ \\
2. Gender: a. Male b. Female c. Non-Binary d. Prefer not to say \\
3. Occupation: a. Student b. Academic/Researcher c. Professional d. Retired e. Unemployed f. Other: (please specify) \\
4. Is Chinese your native language? a. Yes b. No \\
5. Education level: a. High School b. Undergraduate c. Graduate d. Doctorate e. Other: (please specify) \\
6. Vision Condition (select all): a. Normal Vision b. Myopia c. Hyperopia d. Astigmatism e. Presbyopia f. Other: (please specify) \\
7. Do you wear corrective lenses? a. Yes b. No \\
8. Have you been diagnosed with any other eye conditions? a. Yes (please specify: ) b. No \\
9. How often do you read on a mobile device? a. Daily b. Several times a week c. Weekly d. Rarely e. Never \\
10. Daily reading time on mobile devices: a. <10 min b. 10–30 min c. 30–60 min d. 1–2 hrs e. >2 hrs \\
11. Primary devices for reading (select all): a. Smartphone b. Tablet c. Laptop d. E-Reader e. Desktop f. Printed Books g. Other: please specify)  \\
12. Do you often adjust text settings? a. Yes b. No \\
13. If yes, which settings do you adjust most? (Select all): a. Font Size b. Font Style c. Line Spacing d. Brightness e. Color Contrast f. Other: (please specify) \\

\subsection{Daily Survey during Adaptation Period}
\label{appendix: Daily Survey}
1. How many different reading scenarios did you use SituFont in today? a. 3 b. 4 c. 5 d. >5 \\
2. Where did you use SituFont today? (Select all): a. Indoors b. Outdoors c. Public Transport d. Walking e. Stationary \\
3. Lighting conditions? (Select all): a. Bright b. Normal Indoor c. Low Light d. Other: (please specify) \\
4. Did you experience distractions? a. Yes b. No \\
5. If yes, what are the main sources of distraction? (Select all): a. Noise b. Physical Movement c. Notifications d. Visual Distractions e. Other: (please specify) \\
6. Did you experience eye strain or discomfort? a. None b. Slight c. Moderate d. Severe \\
7. Did you manually adjust text settings in SituFont? a. Yes b. No \\
8. If yes, which settings did you adjust? (select all): a. Font Size b. Font Weight c. Line Spacing d. Brightness e. Other: (please specify) \\
9. Did you encounter any problems with SituFont today? a. Yes (please describe: ) b. No \\

\subsection{Post-Test Semi-Structured Interview Protocol}
\label{appendix: Post-test interview}
1. Could you tell me about your overall experience using SituFont over the past week? \\
2. Did you encounter any difficulties adjusting the font settings in SituFont? If so, please describe. \\
3. What did you find most helpful about using SituFont? \\
4. Were there any situations where SituFont did not help or made reading harder? \\
5. How did SituFont compare to the traditional display you used? \\
6. Did you feel you had enough control over SituFont’s text adjustments? \\
7. Would you prefer more options to manually customize text settings? \\
8. Did SituFont ever make adjustments that you disagreed with? If so, which ones? \\
9. Would you continue using SituFont on your device? Why or why not? \\
10. What improvements would you suggest for SituFont? \\

\end{document}